\documentclass[10pt,conference,letterpaper]{sig-alternate}

\usepackage{amsmath,epsfig}
\usepackage{url,cite}
\usepackage{xspace}
\usepackage{colortbl}
\usepackage{subfigure}
\usepackage{dsfont}
\usepackage{boxedminipage}
\usepackage{algorithm}
\usepackage{algorithmic}
\usepackage{amssymb}
\usepackage{times,wasysym}

\begin{document}

\title{AROMA: Automatic Generation of Radio Maps for Localization Systems}

\numberofauthors{3}

\author{
Ahmed Eleryan, Mohamed Elsabagh\\
       \affaddr{Dept. of Computer Engineering}\\
       \affaddr{Faculty of Engineering}\\
       \affaddr{Alexandria University, Egypt}\\
       \email{\{ahmed.eleryan,mohamed.elsabagh\}@alex.edu.eg}\\
\and
Moustafa Youssef\\
       \affaddr{Dept. of Computer Science}\\
       \affaddr{University of Maryland}\\
       \affaddr{College Park, MD 20742, USA}\\
       \email{moustafa@cs.umd.edu}
}

\maketitle

\begin{abstract}
WLAN localization has become an active research field in recent
years. Due to the wide WLAN deployment, WLAN localization provides
ubiquitous coverage and adds to the value of the wireless network by
providing the location of its users without using any additional
hardware. However, WLAN localization systems usually require
constructing a radio map, which is a major barrier of WLAN
localization systems' deployment. The radio map stores information
about the signal strength from different signal strength streams at
selected locations in the site of interest. Typical construction of
a radio map involves measurements and calibrations making it a
tedious and time-consuming operation.

In this paper, we present the design, implementation, and evaluation
of the \textit{AROMA} system that automatically constructs accurate
\textbf{active and passive} radio maps for both device-based and
device-free WLAN localization systems. \textit{AROMA} has three main
goals: high accuracy, low computational requirements, and minimum
user overhead. To achieve high accuracy, \textit{AROMA} uses 3D ray
tracing enhanced with the uniform theory of diffraction (UTD) to
model the electric field behavior and the \textbf{human shadowing
effect}. \textit{AROMA} also automates a number of routine tasks,
such as importing building models and automatic sampling of the area
of interest, to reduce the user's overhead. Finally, \textit{AROMA}
uses a number of optimization techniques to reduce the computational
requirements.

We present our system architecture and describe the details of its
different components that allow \textit{AROMA} to achieve its goals.
We evaluate \textit{AROMA} in two different testbeds. Our
experiments show that the predicted signal strength differs from the
measurements by a maximum average absolute error of 3.18 dBm
achieving a maximum localization error of 2.44m for both the
device-based and device-free cases. Our results also show that
ignoring the effect of the UTD in the device-free case leads to
significant degradation in accuracy up to more than 700\%. We also
relay lessons learned and give directions for future work.
\end{abstract}


\section{Introduction}
WLANs are installed primarily for providing wireless communications.
However, recent research has shown that WLANs can be used in
location determination in indoor environments, without using any
extra hardware \cite{RADAR, ProbWLANLoc, Clustering, Horus, DfPLoc,
Challenges}. Acquiring the location information for a tracked entity
unleashes the possibility of various context-aware applications
including location-aware information retrieval, indoor direction
finding, and intrusion detection.

There are two classes of WLAN location determination systems:
device-based, e.g. \cite{RADAR,Horus} and device-free, e.g.
\cite{Challenges,DfPLoc}.  Device-based systems track the location
of a WLAN-enabled device, such as a laptop or PDA. On the other
hand, device-free systems do not require the entity being tracked to
carry a device and depend on analyzing the effect of the tracked
entity on the signal strength to estimate the entity's position.
Device-free localization systems are composed of a number of access
points (APs) and monitoring points (MPs). The MPs, such as standard
laptops and other wireless-enabled devices, monitor the APs signal
strengths and have fixed locations.

\begin{figure}[!t]
\centering
    \subfigure[Active]{
      \includegraphics[width=0.44\textwidth]{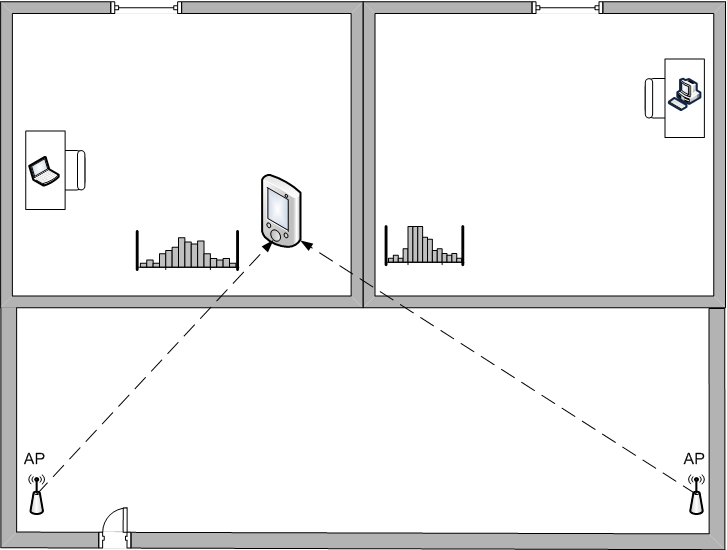}
      \label{fig:rm_active}
    }
    \subfigure[Passive]{
      \includegraphics[width=0.44\textwidth]{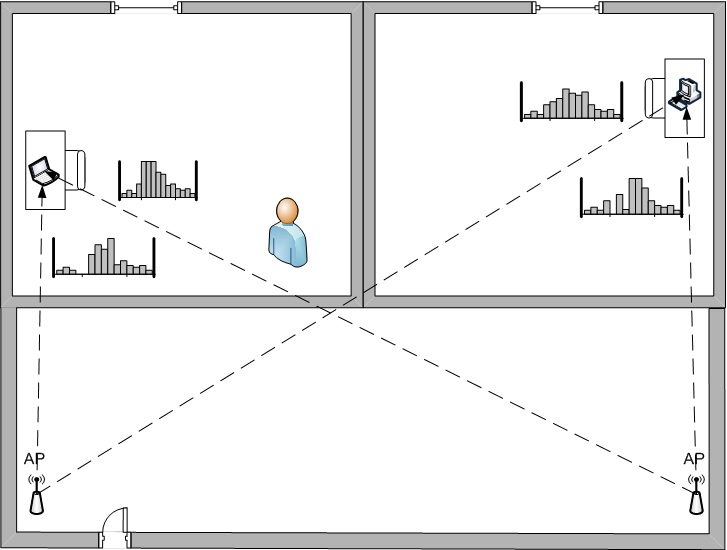}
  \label{fig:rm_passive}
    }
  \caption{Difference between active and passive radio maps construction.
  In a passive radio map, we have a histogram per raw data stream, as compared to a histogram per AP.
  Also, a user does not carry any device when constructing the passive radio map.}
  \label{fig:radio_map}
\end{figure}

Both device-based and device-free systems usually work in two
phases: an offline training phase and an online location
determination phase. During the offline phase, the system collects
signal strengths received from different streams at different
selected locations in the area of interest, and tabulates them into
a so-called radio map. For device-based systems, each stream
represents the signal strength from an AP to the tracked device. For
device-free systems, each stream represents an (AP, MP) pair and the
radio map tabulates the effect of the tracked entity on the fixed
streams. The difference between active and passive radio maps'
construction is illustrated in Figure \ref{fig:radio_map}.

During the location determination phase, the system uses the
information stored in the radio map to estimate the user's location.
Different location determination systems store different information
in the radio map. For example, the Horus system \cite{Horus} stores
the signal strength distribution of the signal strength received
from each AP, while the Radar system \cite{RADAR} stores the average
signal strength received from each AP.

Current methods of radio maps' construction use manual calibrations
making it a tedious and time-consuming operation. Furthermore, each
time the layout of the environment changes or different hardware is
used, the whole process of constructing the radio map has to be
repeated. In addition, the process of radio map construction gets
more complicated, in the device-free case, when the number of
tracked entities increases, since the radio map needs to take all
the combination of the possible tracked entities' locations into
account. For example, for a radio map with $l$ locations and a
system that wants to track up to $n$ entities, the radio map needs
to store information about $l\choose n$ possibilities. This
emphasizes the need for a method to automatically construct the
radio maps for an area of interest.

In this paper we present the \textit{AROMA} (\textbf{A}utomatic
generation of \textbf{R}adi\textbf{O} \textbf{MA}ps) system which
can automatically construct an accurate radio map for a given 3D
area of interest. \textit{AROMA} is unique in supporting automatic
radio map generation for \textbf{both device-based and device-free}
localization systems. To our knowledge, \textit{AROMA} is the first
system to consider radio map generation for device-free systems and
the first to consider \textbf{human effects} in device-based
systems. \textit{AROMA} combines ray tracing with the uniform theory
of diffraction \cite{IntroToUTD} to model both the RF propagation
and human shadowing effect. Ray tracing approximates the
electromagnetic waves as a set of discrete ray tubes that propagate
through the area of interest and that undergo attenuation,
reflection, transmission and diffraction due to the complexity of an
indoor environment. Although ray tracing has been used before in
site-specific radio propagation prediction and several tools have
been developed, e.g. \cite{SeidelRappaport, SiteSpecificThesis,
RTPathGenerator, AppOfRT, RTPropPred}, the main focus of such tools
was the \textbf{radio coverage} problem, i.e. determining the
coverage holes given the APs' positions. This does not require high
accuracy, and therefore, none of these tools account for the human
shadowing effect on the RF signal. Existing papers on the human
shadowing effect, e.g. \cite{UTDTalbi}, illustrate only the theory
behind the modeling and not its application. On the other hand,
propagation modeling for \textbf{localization systems} requires
\textbf{high accuracy}, where variations in the predicted signal
strength can lead to large localization errors. Therefore, the
\textit{AROMA} system has three main goals: (1) to automatically
construct an accurate radio map for a site of interest, (2) to have
efficient computations, and (3) to incur minimum overhead on the
user.

We present the design of the \textit{AROMA} system and give details
about its different components and how they interact to achieve its
goals. We also evaluate the system under two testbeds for the
device-based and device-free cases.

The rest of the paper is organized as follows.
Section~\ref{sec:AROMA} presents the details of the \textit{AROMA}
system. We evaluate the performance of the system under two
different testbeds in Section~\ref{sec:validation}. We discuss our
experience while building the system in Section~\ref{sec:discuss}.
In Section~\ref{sec:related} we discuss related work. Finally,
Section~\ref{sec:conclude} concludes the paper and gives directions
for future work.

\section{The AROMA System}
\label{sec:AROMA}
\subsection{Overview}

\begin{figure}[!t]
    \centering
        \includegraphics[width=0.45\textwidth]{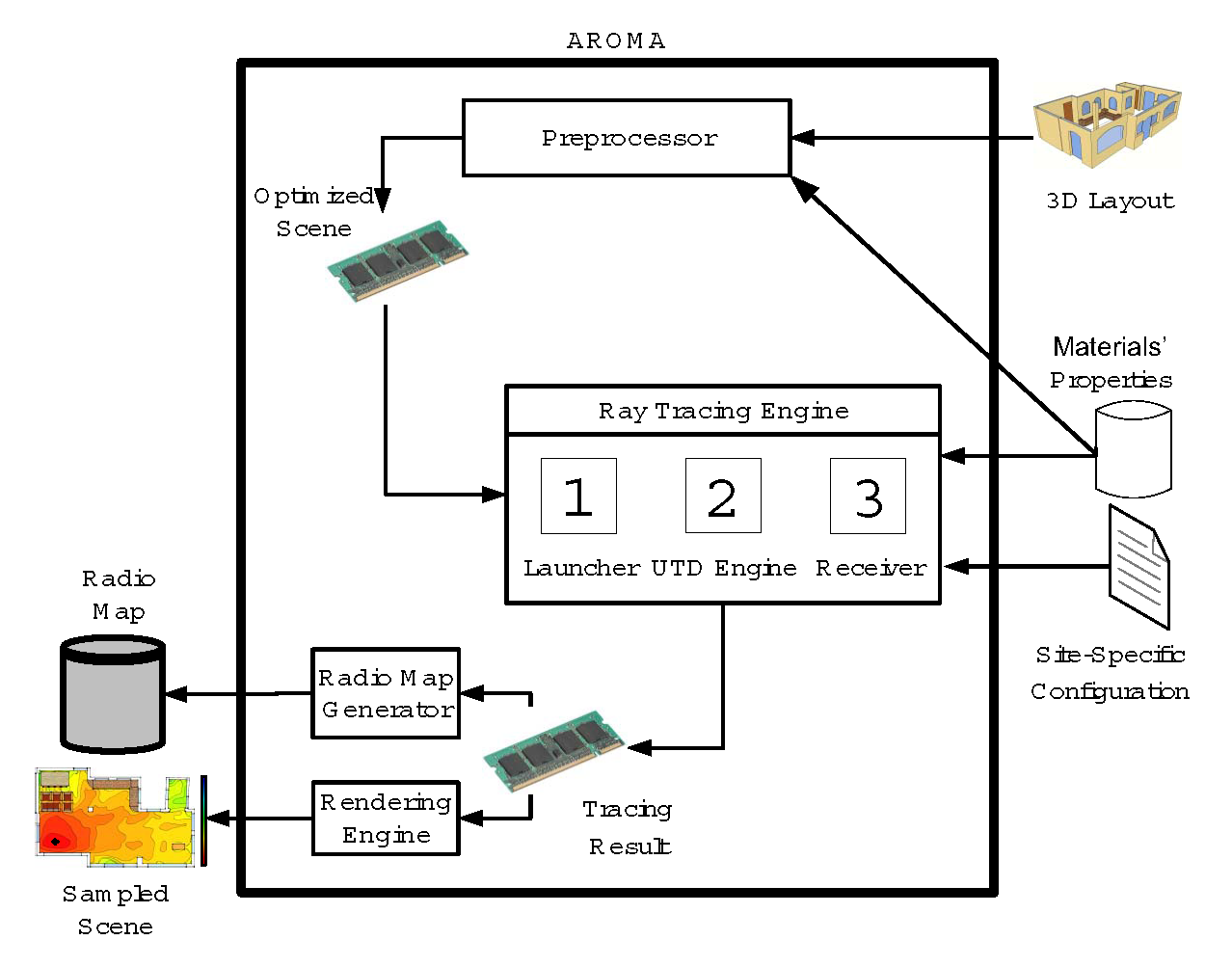}
    \caption{AROMA system architecture.}
    \label{fig:aroma_architecture}
\end{figure}

The \textit{AROMA} System uses site-specific ray tracing, augmented
with the uniform theory of diffraction (UTD), to predict the RF
propagation in a 3D site. Figure \ref{fig:aroma_architecture} shows
the architecture of the \textit{AROMA} System. The input to
\textit{AROMA} is the 3D model of the site of interest which can be
imported automatically from CAD tools or drawn using any free 3D
modeling tool found in the market today such as Google SketchUp and
Blender. The 3D manipulation of the site model is done using the
jMonkeyEngine (jME) \cite{JavaMonkey} which is a high performance
open-source Java-based gaming engine.

Besides the 3D model, the user must provide the site-specific
configuration. This includes the locations and antenna
characteristics of the APs and MPs. The antenna characteristics
include transmitting power, frequency, maximum gain and radiation
pattern. The system comes with two pre-defined antenna radiation
patterns: isotropic and half-wave dipole antenna. The user can also
define a customized radiation pattern depending on the hardware
used. The user has two options to specify the site-specific
configuration:
\begin{enumerate}
    \item Insert them manually using the UI tool.
    \item Provide a configuration file with their locations and antenna characteristics.
\end{enumerate}
The user has similar options when specifying the locations of the
radio map cells.

\textit{AROMA} comes with a built-in DB of the approximate values of
the RF propagation properties of common building materials such as
bricks and concrete. The user has the options of using this DB or
providing customized values using the UI tool.

After providing the 3D model and the site-specific configuration,
the user starts the system. The 3D model is first pre-processed to
extract the edges in the scene. The \textit{Ray Tracing Engine} is
the core of the \textit{AROMA} system and is composed of three
modules: the \emph{Ray Launcher}, the \textit{UTD Engine}, and the
\textit{Ray Receiver}. The \textit{Ray Launcher} samples the
electromagnetic waves emitted from the the APs into a set of
discrete ray tubes covering the area of interest uniformly and each
having an associated electric field. The ray tubes propagate into
the environment and undergo reflections, transmissions and
diffractions. The \textit{Ray Tracing Engine} handles the
interactions of the ray tubes with the environment. The \textit{UTD
Engine} handles the changes in the electric field associated with
the ray tubes resulting from these interactions. The contribution of
each tube in the final received signal strength at the MPs can be
found by the \textit{Ray Receiver}. The tracing result is then
processed by the \textit{Radio Map Generator} to generate the radio
map. A sampled scene with RF prediction levels rendered on its floor
can also be generated by processing the tracing result by the
\textit{Rendering Engine}. The overall algorithm used in the
\emph{AROMA} system is briefed in Algorithm \ref{alg:tracer}. Each
component is explained in details in the next subsections.

\begin{algorithm}[!t]
\caption{Ray Tracer algorithm} \label{alg:tracer}
\begin{algorithmic}
\STATE start preprocessing; \FORALL {access point AP}
  \STATE Queue $Q \leftarrow launchRays()$;
  \FOR {$i = 1$ to depth step 1}
    \WHILE {$\neg empty(Q)$}
      \STATE $ray \leftarrow dequeue(Q)$;
      \STATE $event \leftarrow interaction(ray, site)$;
      \STATE $process(event)$;
      \STATE $utd(event)$;
      \IF {$receivable(receiver, ray)$}
        \STATE $receive(receiver, ray)$;
      \ENDIF
    \ENDWHILE
  \ENDFOR
\ENDFOR \STATE start postprocessing;
\end{algorithmic}
\end{algorithm}

\subsection{Pre-Processing}
The pre-processor works on the 3D model to extract edges in the
model and to assign different materials to the faces in the model.
In addition it constructs data structures that helps in speeding up
the computations.

\subsubsection{Edge Detection}
The 3D model is loaded in the tool as an array of tri-meshes. A
tri-mesh is a set of triangles covering possibly non-contiguous
surfaces. An edge in the model can be spread over multiple
tri-meshes, which need to be merged for accurate modeling of wedge
diffraction and for efficient computations. The edge detection
module uses the hysteresis algorithm \cite{HubeliGross} to detect
the edges in the 3D model. The algorithm consists of two phases.
During the classification phase, each triangle side is assigned a
weight which is the largest angle between the normals to any two
adjacent triangles sharing that side. In the detection phase, a
triangle side is considered an edge if it passes the hysteresis
test, where two thresholds are defined. If the weight of the side is
greater than the upper threshold, then the side is considered an
edge. If it is lower than the lower threshold, the side is
discarded. Other sides are considered edges if they neighbors an
edge.

\subsubsection{Bounding Capsules}
The pre-processor encapsulates edges and wedges by bounding capsules
(a swept sphere containing the object) and assigns them unique IDs
to efficiently test for ray-edge intersection events. Bounding
capsules has an advantage over bounding cylinders as a capsule and
another object intersect if the distance between the capsule's
defining segment and some feature of the other object is smaller
than the capsule's radius \cite{BoundingCapsules}. Bounding capsules
are also used to calculate intersections with the human bodies in
the environment and to efficiently diffract rays around them.

\subsubsection{Materials Detection}
To handle different materials efficiently, the pre-processor assigns
all edges that have the same material the same color. This makes the
3D model more efficient for processing by the jMonkeyEngine.

\subsection{Ray Tracing Engine}
\subsubsection{Ray Launcher}
APs are represented as point sources that emit electromagnetic waves
having spherical wavefronts. These spherical wavefronts are divided
into a number of ray tubes that cover them entirely and have equal
area. Each ray tube is represented by a ray located at its center.
The rays emitted from each AP must experience two forms of
uniformity \cite{DurginRappaport}:
\begin{enumerate}
    \item Large scale uniformity: to guarantee unbiased coverage of rays in the 3D environment,
    \item Small scale uniformity: to guarantee that the angular separation between rays is constant.
\end{enumerate}

\begin{figure}[!t]
    \centering
        \includegraphics[width=0.5\textwidth]{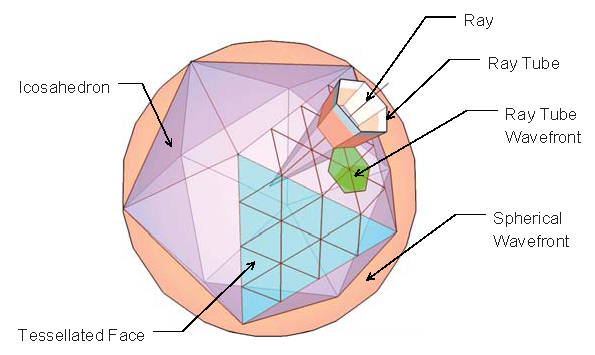}
    \caption{Ray launching by tessellating the faces of an icosahedron. The transmitter
    is located at the center of the icosahedron. Each icosahedron face is tessellated into
    smaller triangles using a tessellation frequency (N). Rays are launched passing through
    the vertices of the produced triangles.}
    \label{fig:ray_launcher}
\end{figure}

These conditions are satisfied by emitting rays through the vertices
of an icosahedron whose center is the transmitter. An icosahedron
has 12 vertices and the angular separation between rays emitted from
its vertices equals $69^\circ$. To achieve a better angular
resolution, the face of the icosahedron is divided into smaller
triangles using a tessellation frequency N \cite{SeidelRappaport}.
The rays are then emitted from the vertices of the formed triangles
and the ray tubes are hexagonal in shape as shown in Figure
\ref{fig:ray_launcher}.
A good approximation for the angular separation between rays in
this case is \cite{DurginRappaport}:
\begin{equation}
    \alpha =\frac{69.0{}^\circ }{N}
    \label{eq:angular_separation}
\end{equation}

Section \ref{subsec:ray_tracing_parameters} discusses the values used for the ray tracing parameters.

\subsubsection{Ray Tracer}

The complex indoor environment causes the ray tubes to change their
original direction through either reflection, transmission, or
diffraction. This results in a phenomena known as multipath fading
where the transmitted signal reaches a receiver via multiple paths.

\begin{figure}[!t]
    \centering
        \includegraphics[width=0.5\textwidth]{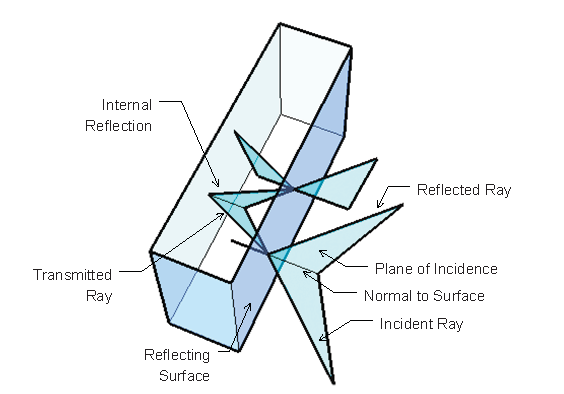}
    \caption{Reflection and transmission of rays. A ray incident on a surface experiences reflection and
    transmission. The transmitted ray suffers also from internal reflection and transmission.}
    \label{fig:reflection_transmission}
\end{figure}

Each ray is traced for different kinds of interactions with the 3D
site. A ray incident on an object produces a reflected and a
transmitted rays as shown in Figure
\ref{fig:reflection_transmission}. Reflected rays satisfy the laws
of reflection:
\begin{enumerate}
    \item The incident ray, reflected ray, and the normal to the reflecting surface are coplanar.
    \item Angle of incidence equals the angle of reflection.
\end{enumerate}

And the transmitted rays satisfy the laws of refraction:
\begin{enumerate}
    \item The incident ray, refracted ray, and the normal to the reflecting surface are coplanar.
    \item The ratio of the sines of the angles of incidence and refraction is equivalent to the opposite ratio of the indices of refraction (Snell's Law).
    \begin{equation}
        \frac{\sin\theta{}_{1}}{\sin\theta{}_{2}} = \frac{n{}_{2}}{n{}_{1}}
        \label{eq:snell_law}
    \end{equation}

\end{enumerate}

Where $\theta{}_{1}$ is the angle of incidence, $\theta{}_{2}$ is
the angle of refraction, $n{}_{1}$ and $n{}_{2}$ are the refractive
indices of the first and second mediums respectively.

\begin{figure}[!t]
    \centering
        \includegraphics[width=0.35\textwidth]{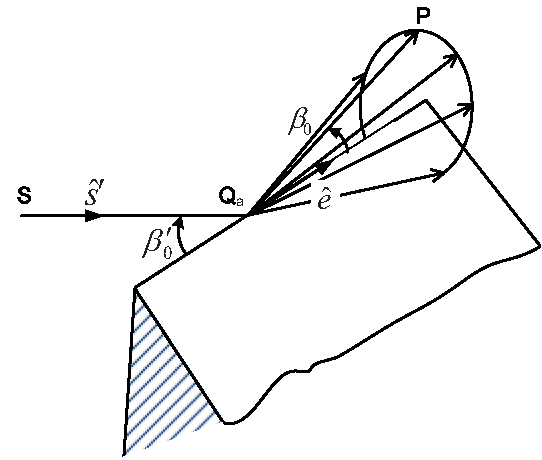}
    \caption{Wedge diffracion. A ray intersecting with an edge produces a set of diffracted rays
    forming a diffraction cone. $\beta_0$ is the angle between the incident ray and the edge,
        ${\hat{s}}^{'}$ is the direction of the incident ray, $\hat{e}$ is
        the direction of the edge, and $\hat{s}$ is the direction of the
        diffracted ray. ${Q_a}$ is the point of intersection, and $P$ is any point in the space
        lying on the produced cone.}
    \label{fig:wedge_diffraction}
\end{figure}

A ray incident on an edge produces a set of diffracted rays forming
what is called a diffraction cone (as shown in Figure
\ref{fig:wedge_diffraction}) obeying the law of diffraction:

\begin{quotation}
A diffracted ray and the corresponding incident ray make equal angles with the edge at the point of diffraction, and they lie on opposite sides of the plane normal to the edge at the point of diffraction.
\end{quotation}
\begin{equation}
\sin{\beta}_0 = | {\hat{s}}^{'} \times \hat{e} | = | \hat{s} \times \hat{e} |
\label{eq:law_diffraction}
\end{equation}

Where $\beta_0$ is the angle between the incident ray and the edge,
${\hat{s}}^{'}$ is the direction of the incident ray, $\hat{e}$ is
the direction of the edge, and $\hat{s}$ is the direction of the
diffracted ray.

Each time the ray makes an interaction with the environment, its
depth is incremented. The tracing of a ray ends in one of two cases:

\begin{enumerate}
    \item The depth reaches a maximum user-defined value.
    \item The power associated with the ray decreases below a defined minimum value.
\end{enumerate}

\subsubsection{UTD Engine}

Electromagnetic waves are discretized using Geometrical Optics (GO)
into a set of ray tubes. Each ray tube propagates in a direction
identified by the ray located at its center, and has associated
electric and magnetic fields orthogonal to each other and to the
direction of propagation. GO can account for reflection and
transmission \cite{IntroToUTD}. The reflected and transmitted
electric fields are related to the incident electric field using
Fresnel's Field Coefficients \cite{BornWolf}. However, GO fails to
account for the electric fields in the shadow regions which occur
due to electric field diffraction. UTD addressed the GO deficiency
which predicts zero electric field in human shadowed regions, and
thus UTD is used in modeling wedge diffraction. Another reason for
using UTD is that it overcomes the drawbacks of the Geometrical
Theory of Diffraction (GTD), which is an extension of geometrical
optics that accounts for diffraction \cite{KELLER}. GTD suffers from
some problems \cite{IntroToUTD}, the most serious of them is that
GTD predicts singular electric fields near the transition regions.
In UTD, The diffracted electric field is related to the incident
electric field using UTD diffraction coefficients \cite{IntroToUTD}.

The construction of device-free radio maps requires modeling of
human's body effect on RF signals.
At microwave frequencies and higher, the human body constitutes an
impassable reflector for electromagnetic waves. That is, incident
waves are reflected and diffracted off the body, along other
interactions with the surrounding environment. Previous work in
human modeling has shown a strong correlation between the RF
characteristics of the human body and a metallic circular cylinder
\cite{UTDTalbi} in indoor radio channels. Therefore, we use a
metallic cylinder  to model the human body with radius $0.15m$, and
height $2m$ \cite{UTDTalbi}.

All equations related to the UTD Engine are
illustrated in Appendix \ref{appendix:A}.

\subsubsection{Ray Receiver}

\begin{figure}[!t]
    \centering
        \includegraphics[width=0.45\textwidth]{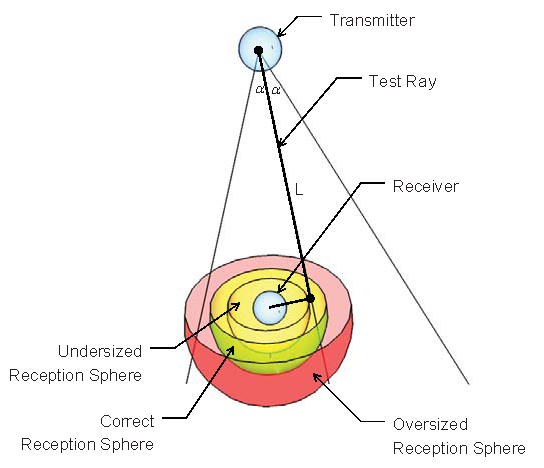}
    \caption{Reception sphere. The size of the reception sphere should be large enough
    to intersect only one ray from a given wavefront.}
    \label{fig:reception_sphere}
\end{figure}

MPs are represented as point sinks. A ray is received if the MP lies
within its ray tube. Alternatively, and for efficiency, we can
consider a ray received if it lies within a certain distance from
the MP. This distance depends on the size of the ray tube at the MP.
The reception sphere model is based on this observation
\cite{SeidelRappaport}. A sphere is formed around the MP such that
at most one ray from a wavefront intersects with the sphere, as
illustrated in Figure \ref{fig:reception_sphere}. The reception
sphere radius equals to the radius of the circle circumscribed about
the hexagonal wavefront of the ray tube. Different rays have
different reception sphere radii calculated as:

\begin{equation}
    r=\ \frac{\alpha \ l}{\sqrt{3}}
    \label{eq:reception_sphere_radius}
\end{equation}

Where $\alpha$ is the angular separation between rays and $l$ is the
total unfolded distance traveled by the ray till the perpendicular
projection of the receiver on the ray path.

\subsection{Post-Processing}

\subsubsection{Radio Map Generator}
A radio map is constructed by dividing the area of interest into a
number of locations, each location has a corresponding cell in the
radio map. The user should provide a list of the radio map cell
locations and the required type of radio map, i.e. active or passive
radio map (Figure \ref{fig:radio_map}). As another option, the user
can select a radio map spacing and the system can automatically
generate the radio map locations.

\subsubsection{Rendering Engine}
\begin{figure}[!t]
    \centering
        \includegraphics[width=0.45\textwidth]{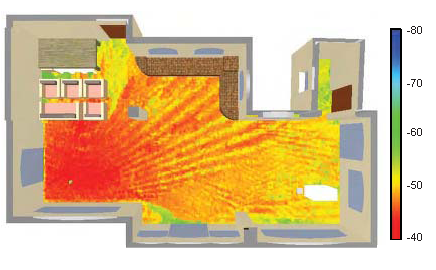}
    \caption{An example of the output of the rendering engine. At the sampling step, an isotropic
antenna is virtually positioned at each sampling point and the
overall sampling result is then bi-cubically interpolated over the
whole floor area.}
    \label{fig:sampling}
\end{figure}

The tracer can sample the site area to predict the signal strength
across the entire site of interest. The tracing result are then
color-shaded and layered over the site floor by the
\textit{Rendering Engine}. At the sampling step, an isotropic
antenna is virtually positioned at each sampling point and the
overall sampling result is then bi-cubically interpolated over the
whole floor area (Figure \ref{fig:sampling}).

\section{System Evaluation}
\label{sec:validation}

Our validation is composed of two experiments: one for the
device-based radio map generation and the other for the device-free
radio map generation. We used two Cisco Aironet 1130G Series 802.11G
Access Point and two D-Link DWL-G650 NICs.
The system is currently implemented in the Windows OS. A NIC
Query\cite{NicQuery} driver that provides an API for user-level
queries of NDIS\cite{NDISProt} devices is used to collect the signal
strength samples. We collected 60 samples for each location in each
experiment. Each experiment has different configurations that will
be illustrated in details in the
next subsections.

\subsection{Device-based Radio Map Generation}
\begin{figure}[!t]
    \centering
        \includegraphics[width=0.5\textwidth]{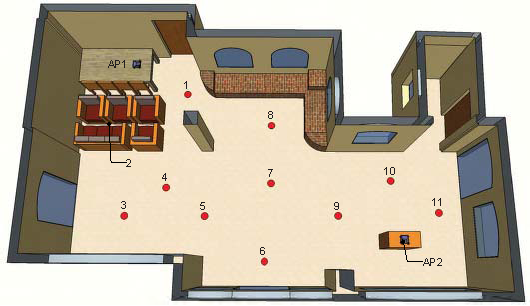}
    \caption{Device-based experiment layout. The figure highlights the locations of APs and radio map locations.}
    \label{fig:device_based_layout}
\end{figure}

The experiment was conducted in a typical apartment with an area of
$700{ft}^2$. The environment contains furniture and is composed of
different materials, like bricks, concrete, wood and glass. We
marked 11 different locations covering the entire area as shown in
Figure \ref{fig:device_based_layout}. The APs' configurations are
summarized in Table \ref{tab:APsConfigurations}.

\begin{table}[!t]
    \centering
        \begin{tabular} {|p{1.3in}|p{0.8in}|}
        \hline $P_t$ & $2 mW$\\
            \hline $Antenna gain (G_{max})$ & $3.0 dBi$ \\
            \hline $Frequency (f)$ & $2.4 GHz$ \\
            \hline Antenna Type & Isotropic \\
            \hline
        \end{tabular}
    \caption{APs' configurations for the device-based experiment.}
    \label{tab:APsConfigurations}
\end{table}

\begin{figure}[!t]
  \begin{center}
    \subfigure[RSS from AP1] {
        \label{fig:device_based_ap1}
        \includegraphics[width=0.5\textwidth]{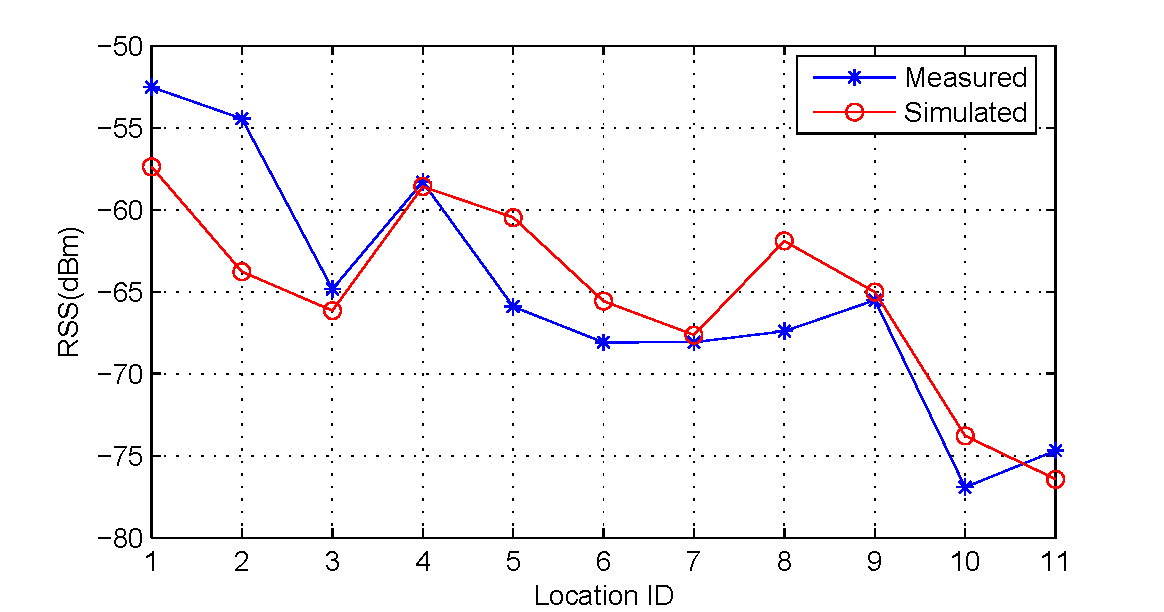}
    }
    \subfigure[RSS from AP2] {
        \label{fig:device_based_ap2}
        \includegraphics[width=0.5\textwidth]{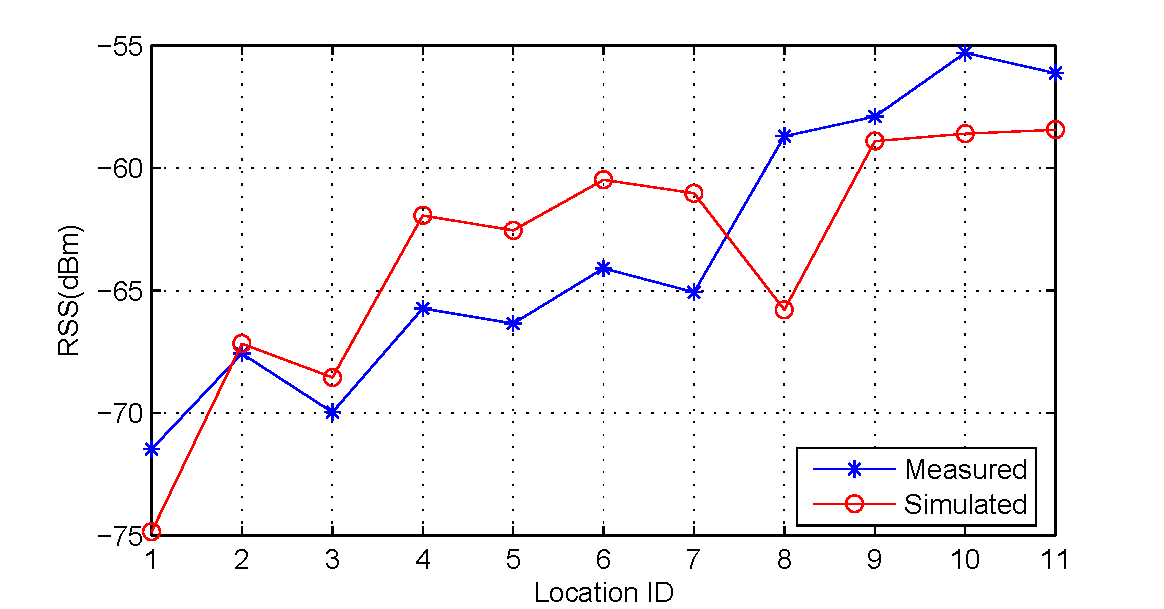}
    }
  \end{center}
  \caption{Device-based experiment, measured vs. simulated RSS. Location ID refers to the IDs in Figure \ref{fig:device_based_layout}.}
  \label{fig:device_based_ap1_ap2}
\end{figure}

Figure \ref{fig:device_based_ap1_ap2} shows the simulated and
measured values for the two APs. Table \ref{tab:db_validation}
summarizes the results. The figure shows that the tool gives the
same trend as the measurements with a maximum average absolute error
of less than 3.2 dBm for the two APs. This shows that the tool can
be used to model more complex scenarios of behavior in a
device-based setting.

\subsection{Device-free Radio Map Generation}

\begin{figure}[!t]
    \centering
        \includegraphics[width=0.5\textwidth]{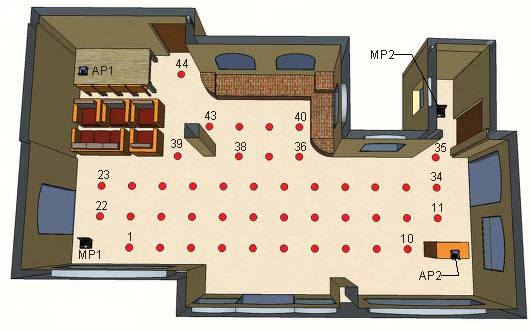}
    \caption{Device-free experiment layout. The figure highlights the locations of APs, MPs, and radio map locations.}
    \label{fig:device_free_2_layout}
\end{figure}

The experiment was held in the same environment as that of the
device-based radio map experiment. 44 locations were chosen and are
illustrated in Figure \ref{fig:device_free_2_layout}. A new location
is introduced here, namely location 0, which represents the
environment without the human. The APs' configurations are
summarized in Table \ref{tab:APsConfigurations}.

\begin{figure}[!t]
    \begin{center}
      \subfigure[RSS from AP1 by MP1] {
        \label{fig:df_44_ap1_mp1}
        \includegraphics[width=0.5\textwidth]{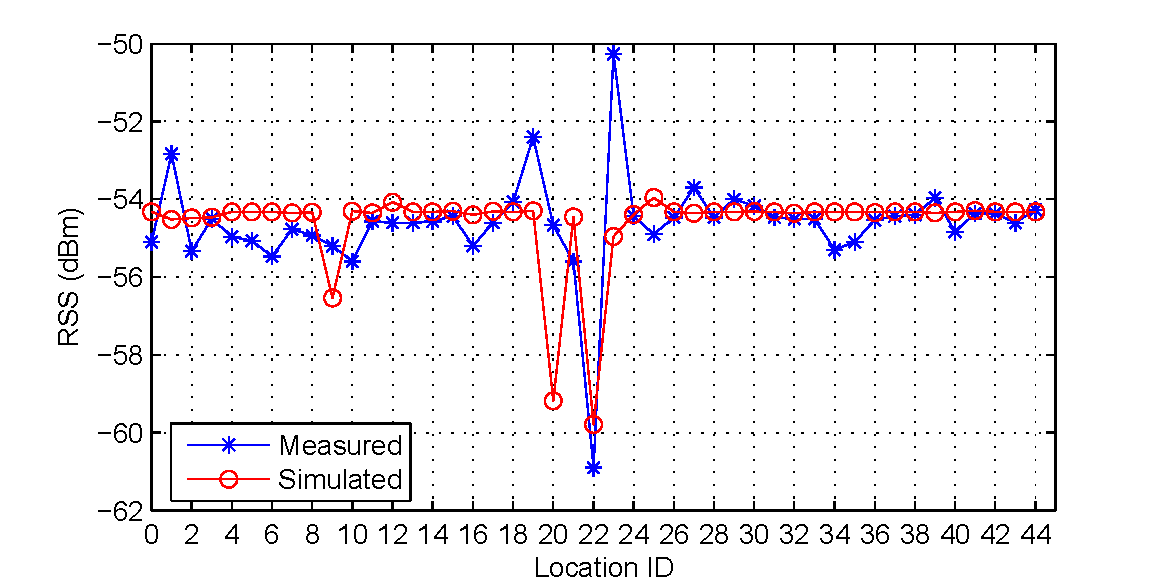}
    }
    \subfigure[RSS from AP1 by MP2] {
        \label{fig:df_44_ap1_mp2}
        \includegraphics[width=0.5\textwidth]{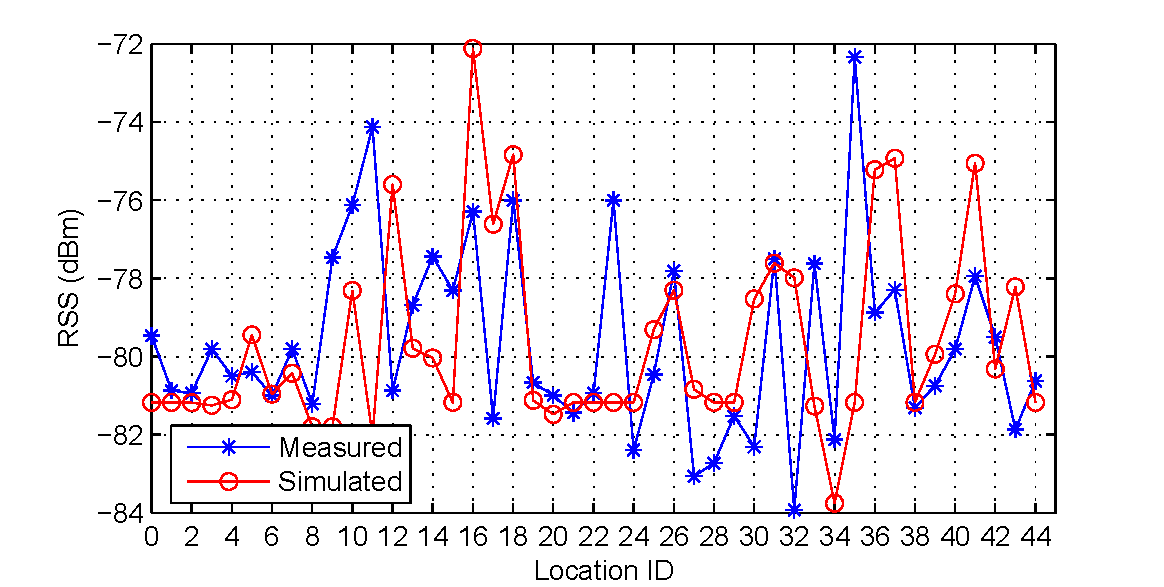}
    }
    \subfigure[RSS from AP2 by MP1] {
        \label{fig:df_44_ap2_mp1}
        \includegraphics[width=0.5\textwidth]{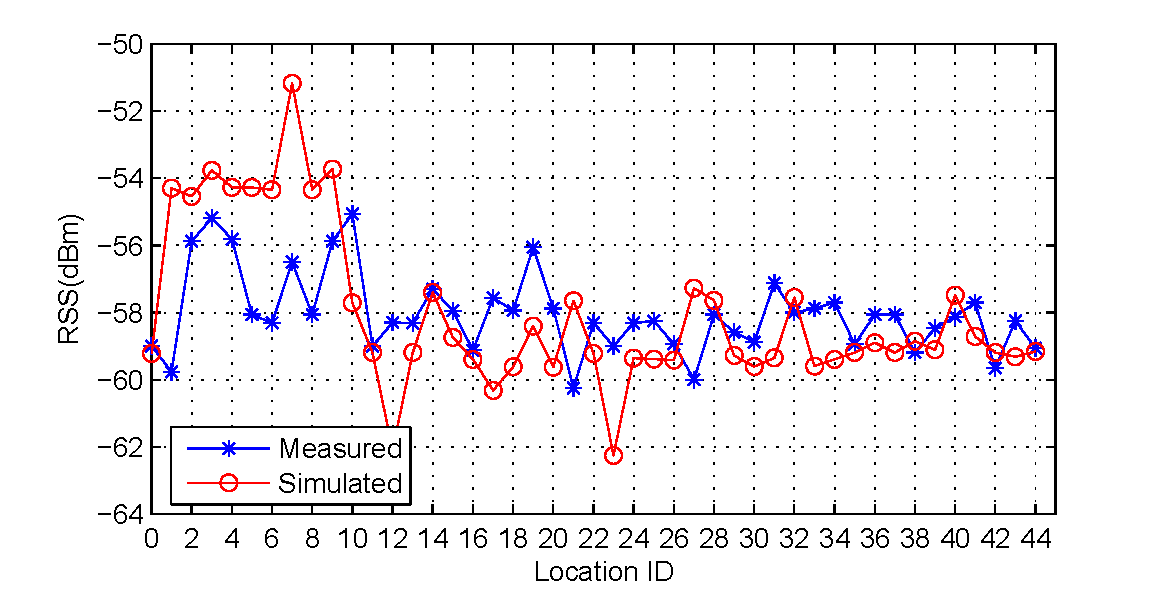}
    }
        \subfigure[RSS from AP2 by MP2] {
        \label{fig:df_44_ap2_mp2}
        \includegraphics[width=0.5\textwidth]{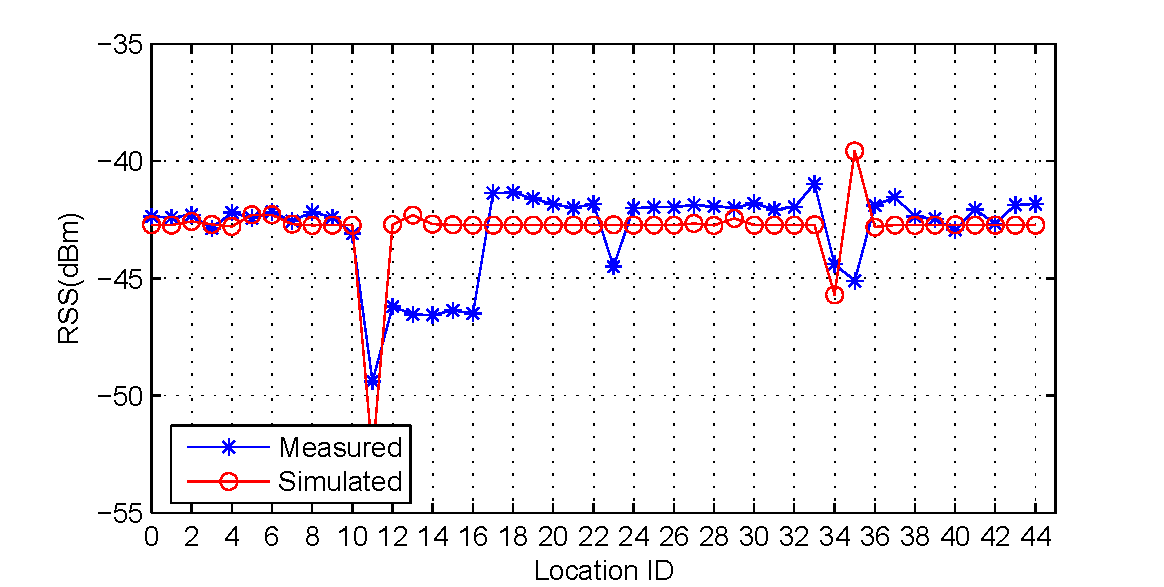}
    }
    \end{center}
\caption{Device-free experiment, measured vs. simulated RSS.
Location ID refers to the IDs in Figure
\ref{fig:device_free_2_layout}.} \label{fig:df_44_results}
\end{figure}

Figure \ref{fig:df_44_results} shows the simulated versus measured
RSS for the experiment. The results are summarized in Table
\ref{tab:df_2_validation}. The results show that simulated values
are close to the measured values with a maximum average absolute
error of 2.17 dBm for all streams. The figure also shows that the
human effect is maximum when the person is cutting the
line-of-sight, e.g. locations 11, 34, and 35 in Figure
\ref{fig:df_44_ap2_mp2}. This is captured by both the tool and the
actual measurements which shows that the tool can be used to model
more complex scenarios of behavior in a device-free setting.

\begin{table}[!t]
    \centering
    \begin{tabular} {|p{1.5in}|p{0.6in}|p{0.6in}|}
    \hline
            & AP1 & AP2 \\
    \hline
                RMSE &  4.00 dBm& 3.4 dBm\\
                Average absolute error & 3.2 dBm& 3.1 dBm\\
                Standard deviation & 3.18 & 3.10 \\
        \hline
    \end{tabular}
    \caption{Device-based experiment, measured vs. simulated results.}
    \label{tab:db_validation}
\end{table}

\begin{table}[!t]
    \centering
                \begin{tabular} {|p{.3in}|p{0.85in}|p{0.65in}|p{0.65in}|}
                \hline
                        \multicolumn{2}{|p{1in}|}{} & AP1 & AP2 \\
                \hline
                            MP1 & RMSE & 1.19 dBm& 2.11 dBm\\
                            & Avg. Abs. Err. & 0.71 dBm& 1.61 dBm\\
                            & Stdev & 1.2 & 2.13 \\
                                \hline
                            MP2& RMSE & 3.02 & 1.75 \\
                             & Avg. Abs. Err. & 2.17 dBm& 1.21 dBm\\
                               & Stdev & 3.06 & 1.77 \\
                 \hline
                \end{tabular}
        \caption{Device-free experiment, measured vs. simulated results.}
        \label{tab:df_2_validation}
\end{table}

\subsection{Localization Performance}
In this section, we compare the performance of a simple
nearest-neighbor localization classifier, similar to the Radar
system \cite{RADAR}, when trained by measured and simulated data. We
divide the collected samples into two parts with a ratio of 2 to 1.
The larger part is used as a training data for the
measurements-based classifier and the other as a testing data for
both types of classifiers.

The results are summarized in Table \ref{tab:localization_perf}. For
the device-based localization system, the mean distance error
calculated for the measurement-based and simulation-based
classifiers are 0.89m and 1.61m respectively. Similarly, for the
device-free case, the mean distance error calculated for the
measurement-based and simulation-based classifiers are 0.96m and
2.44m respectively. Note that since cross-validation is used to
evaluate the measurement-based classifier, its accuracy is over
estimated. Using an independent test set, the accuracy of the
measurement-based classifier will be worse, while the accuracy of
the simulation-based classifier will not be affected. Therefore, the
difference between the measurement-based and simulation-based
classifiers will be less.

\begin{table}[!t]
    \centering
        \begin{tabular} {|p{0.8in}|p{0.9in}|p{0.8in}|}
        \hline Radio map & Measurements-based Classifier & Simulation-based Classifier \\
            \hline Device-based & $0.89m$ & $1.61m$\\
            \hline Device-free & $0.96m$ & $2.44m$\\
            \hline
        \end{tabular}
    \caption{Localization performance (mean distance error) for the device-based and device-free experiments.}
    \label{tab:localization_perf}
\end{table}

\subsection{Effect of Ignoring the UTD}
\label{sec:NO_UTD} 
\begin{figure}[!t]
    \begin{center}
      \subfigure[RSS from AP1 by MP1] {
        \label{fig:df_44_nutd_ap1_mp1}
        \includegraphics[width=0.5\textwidth]{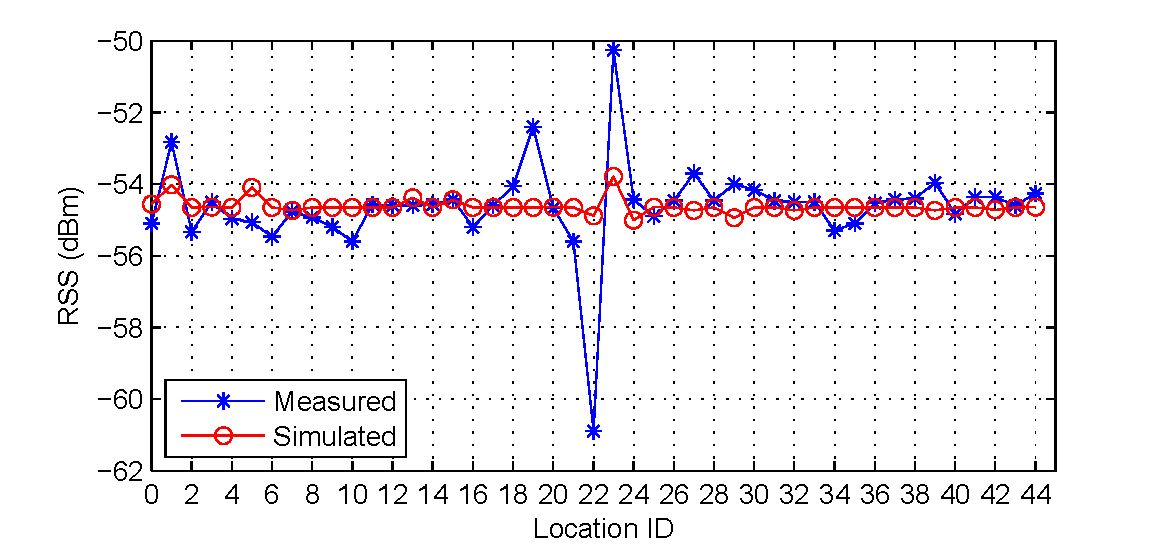}
    }
    \subfigure[RSS from AP1 by MP2] {
        \label{fig:df_44_nutd_ap1_mp2}
        \includegraphics[width=0.5\textwidth]{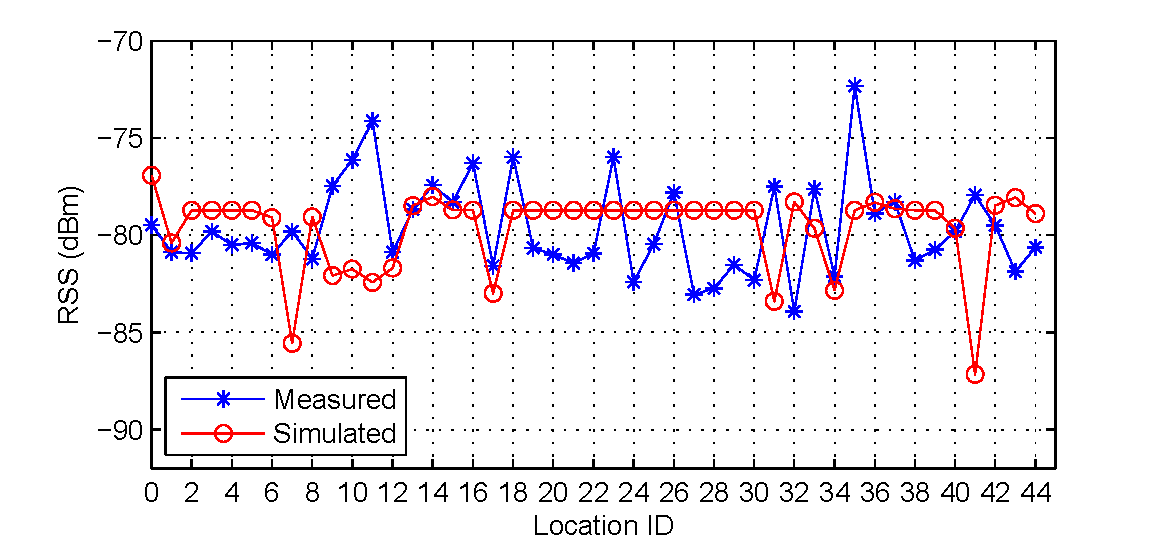}
    }
    \subfigure[RSS from AP2 by MP1] {
        \label{fig:df_44_nutd_ap2_mp1}
        \includegraphics[width=0.5\textwidth]{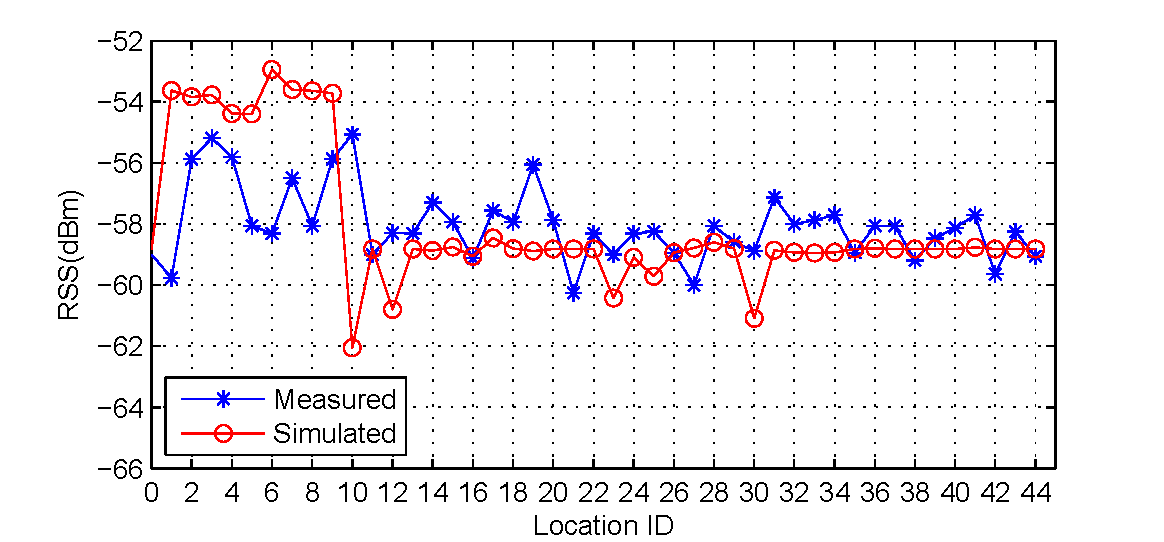}
    }
        \subfigure[RSS from AP2 by MP2] {
        \label{fig:df_44_nutd_ap2_mp2}
        \includegraphics[width=0.5\textwidth]{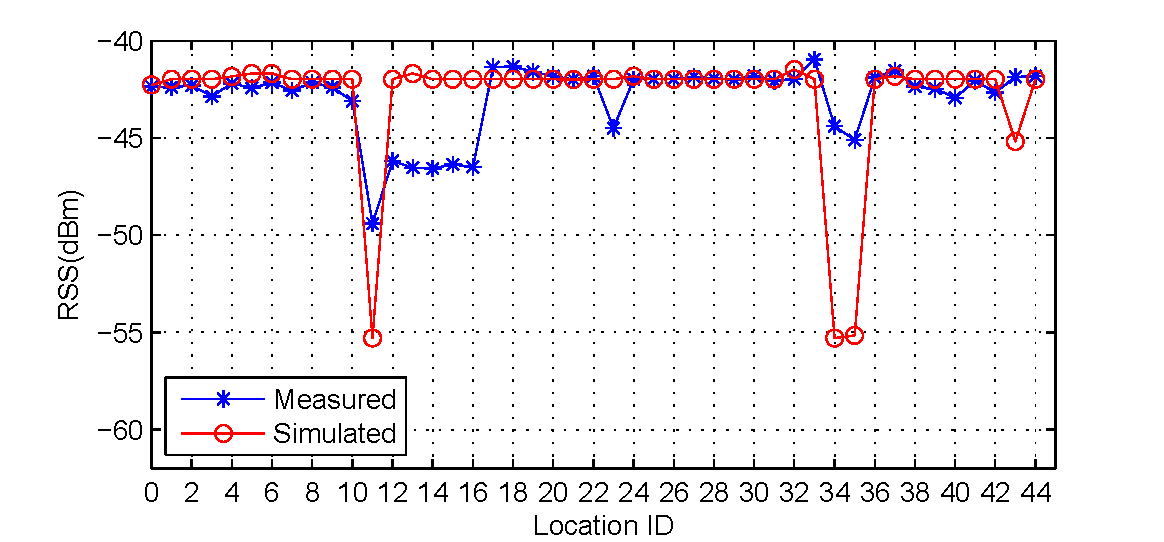}
    }
    \end{center}
\caption{Device-free experiment when the UTD engine is not used,
measured vs. simulated RSS.} \label{fig:df_NO_UTD_results}
\end{figure}

In this section, we show the effect of not using
the UTD engine, which is the common practice in previous work in the
area of RF propagation. Turning off the UTD engine disables all the
interaction with the human except for the attenuation effect, i.e.
the signal is attenuated by a constant amount if the human is
obstructing the signal. Figure \ref{fig:df_NO_UTD_results} shows the
results and Table \ref{tab:df_NO_UTD_validation} summarizes them.
The results show that ignoring the contribution of the UTD engine
degrades the performance significantly, up to 706\%. This is
specially true for MP2, where its location makes it significantly
affected by the UTD effects (Figure \ref{fig:device_free_2_layout}).

\section{Discussion}
\label{sec:discuss} In this section, we discuss different aspects of
the \textit{AROMA} system and our experience while developing it.

\subsection{Computational Efficiency}
\label{subsec:comp_eff} The \textit{AROMA} system uses a number of
techniques to reduce the computational overhead of ray tracing.
Using bounding capsules, combining edges, and mapping materials to
colors are all performed at the pre-processing stage to increase the
efficiency of different modules while the ray tracing engine is
running. In addition, the ray launcher's parameters can be tuned for
maximum accuracy and performance. Finally, the concept of reception
sphere is used in the ray receiver to further reduce computational
requirements.

\subsection{Ease of Use}
\label{subsec:comp_eff} Our \textit{AROMA} system contains a number
of modules that enhance the user experience and automate common
tasks. The UI module allows the user to use point-and-click to enter
the locations of APs and MPs. Different wizards are provided for the
user to enter the APs and MPs parameters, e.g. the antenna's
pattern, and materials properties. A database of different
materials' properties, along with a set of predefined antenna
patterns are also included to reduce the overhead of entering this
information. In addition, the user can configure the system to
automatically select the radio map locations.

\subsection{Prediction Precision}
\label{subsec:prediction_precision} In a perfect world, the measured
RSS should match the predicted RSS perfectly. There are different
reasons for deviation of the predicted values from the measured
ones. Extensive work has been done in the characterization of signal
loss for various materials in indoor environments. However, in
practice the physical and electrical characteristics of materials
vary considerably from theory. This variance affects greatly the
accuracy of signal prediction. In a similar manner, the human body
is made up of water that causes the signal to attenuate. The amount
of signal loss depends on the size of a human body and its
orientation that varies considerably from one person to another.

In addition, there are always discrepancies between hardware
characteristics and the characteristics included in its data sheet.
The effect of these discrepancies is inevitable. The most
influencing discrepancy is that resulting from the antennas
radiation patterns and orientation. The difference between where the
radio point location is selected and where the human is actually
standing is another reason for deviation. In summary, the reasons of
deviation of predicted values from the measurements include:
\begin{itemize}
    \item Lacking of a perfect 3D model of the environment.
    \item Lacking of the precise electrical characteristics of the materials in the environment.
    \item Hardware discrepancies.
    \item UTD assumes diffraction off perfect conductors, yet the environment contains diverse dielectrics.
    \item Surrounding noise and imprecise human location.
\end{itemize}

\begin{table}[!t]
    \centering
                \begin{tabular} {|p{.3in}|p{0.85in}|p{0.65in}|p{0.65in}|}
                \hline
                        \multicolumn{2}{|p{1in}|}{} & AP1 & AP2 \\
                \hline
                            MP1 & RMSE &  2.87 dBm& 5.91 dBm\\
                            & Avg. Abs. Err. & 2.60 dBm & 5.49 dBm\\
                            & \% degradation & 266\%& 241\%\\
                            & Stdev &  1.22 & 2.20 \\
                                \hline
                            MP2& RMSE & 11.53 dBm& 10.19 dBm\\
                             & Avg. Abs. Err. & 11.01 dBm&  9.76 dBm\\
                            & \% degradation & 407\%&  706\%\\
                               & Stdev & 3.46 & 2.95 \\
                 \hline
                \end{tabular}
        \caption{Device-free experiment when the UTD engine is not used, measured vs. simulated results.}
        \label{tab:df_NO_UTD_validation}
\end{table}

\subsection{Ray Tracing Parameters}
\label{subsec:ray_tracing_parameters}

The ray tracing algorithm has two parameters namely: tessellation
frequency of APs ($N$) and tracing depth ($d$). Throughout our
experiments, we noticed that the results are not sensitive to the
tracing depth. The reason behind this is that rays at higher depths
have smaller power and thus make little contribution to the received
signal. We used a tracing depth of 4 in our experiment. A
tessellation frequency between 4 and 8 gives both good accuracy and
efficient computations.

\subsection{Radio Maps for Probabilistic Localization Systems}
The localization results reported in this paper are for
deterministic location determination systems that depend on storing
the average RSS in the radio map. Probabilistic location
determination systems, e.g. \cite{Horus}, store RSS distributions in
the radio map. In order to generate a probability distribution,
different approaches can be used: Perturbing the location of the
device or human, for device-based and device-free systems
respectively, leads to changing the generated radio map and hence
can be used to estimate the probability distribution. Similarly, the
locations and antenna patterns of APs and MPs and the state of doors
and windows can be changed to achieve a similar effect. Another
possibility is to develop models for the relation between the
average RSS of a stream and its variance. Previous studies have
reported that the higher the average RSS, the higher the variance
\cite{Horus}. This effect can be more accurately quantified and
used.

\subsection{Dynamic Radio Map Generation}
The radio map can change also from time to time based on the
environment conditions. To capture these changes, the antennas'
parameters and the building materials can be adjusted based on
environment changes, such as temperature and/or humidity over the
day or for a given semester in the year. Other effects that are
dependent on the time of the day, such as human density, can also be
included in the tool, e.g. by randomly placing humans in the area of
interest based on the time of day.

Feeding the tool with a small number of tuning measurements from the
environment can be used as another technique to change the system's
parameters to capture the changes in the environment.

\subsection{Human Modeling}
Currently, the \textit{AROMA} system models the human as a perfectly
conducting cylinder \cite{UTDTalbi}. Although this gives good
results, we believe that the system's accuracy can be enhanced,
especially for the device-free case, by having a better human model.
The main drawback of the cylinder model is that it does not capture
the effect of the user's orientation. This is an area for ongoing
work.

\subsection{Other Uses}
In this paper, we showed how the \textit{AROMA} system can be used
in constructing accurate radio maps for device-based and device-free
localization systems. However, the \textit{AROMA} system can also be
used in a number of different applications. In addition to the
traditional site planning, such as determining the placement of APs
to maximize coverage and detect coverage holes, \textit{AROMA} can
be used to select the locations of APs and MPs for maximum combined
coverage and localization accuracy. \textit{AROMA} can be also used
in the design of new algorithms for localization systems, especially
for the challenging and open research area of device-free tracking
and identification \cite{Challenges}. The system can be used to test
the effect of the existence of different humans in the area on the
constructed radio map and the effect of different instance of the
same object. Another application for the system is in studying the
effect of the new smart APs that have automatic power adjustment on
the performance of localization systems.

\section{Related Work}
\label{sec:related}

Indoor radio propagation has been an active field of research
\cite{SeidelRappaport, SiteSpecificThesis, RTPathGenerator, AppOfRT,
RTPropPred}, especially for cellular networks. Seidel and Rappaport
discuss different radio propagation models in
\cite{SeidelRappaport}. Based on the ray-tracing technique, most
authors model the propagation loss using the \textit{simple
attenuation model}:
\begin{equation}
P\left(d\right) = P\left(d_0\right) - 10 \times n \times
{\log10}\left(\frac{d}{d_0}\right) \label{eq:attenuation_exponent}
\end{equation}

Where $P\left(d\right)$ is the power at distance $d$ from the
transmitter, $P\left(d_0\right)$ is the power at a reference
distance, and $n (\geq 2) $ is the attenuation exponent.

The simple attenuation model requires extensive measurements in
order to estimate a best-fit attenuation exponent that reduces the
overall error for a \textit{certain} site. In \cite{Jenkins}, Hills
et al. report that an attenuation exponent of 2.60 results in the
best fit in the buildings on the Carnegie Mellon University campus.
Beside requiring extensive site-specific measurements to estimate
its parameters, the model does not account for attenuation effects
from typical environments, like doors, walls, and other different
materials.

A more sophisticated model is the \textit{partition model}. The
partition model takes into consideration path loss caused by indoor
partitions, like walls and doors. This simple model cannot give
sufficient accuracy for localization systems. For example, the Radar
system, which accounts for multiple wall attenuation effects along
the direct path between transmitters and receivers, has a mean error
of 3.4m \cite{RADAR}. As we show in Section \ref{sec:validation} and
Table \ref{tab:localization_perf}, the \textit{AROMA} system can
give accuracy of 1.61 m in a similar environment. In addition, the
proposed tools has a number of other uses as discussed in Section
\ref{sec:discuss}.

Another model that is closely related to the partition model is the
\textit{site-specific model}. The site-specific model behaves like
the partition model, but it needs more site-specific parameters such
as materials and geometrics. Ali et al. \cite{HassanAli} introduced
a novel model that relate the average path power to the
site-specific parameters. However, such systems do not consider the
human effect and their goal is to study coverage and not
localization.

In \cite{ARIADNE}, Biaz et al. presented ARIADNE, a dynamic
device-based indoor radio map construction and device-based
localization system. given a number of actual measurements, the
system adapts to temporal changes of radio propagation. Global
attenuation parameters are automatically estimated using simulated
annealing search. The system requires site-specific measurements and
does not handle the human effect or the device-free case. The
results in Section \ref{sec:NO_UTD} show that ignoring the human
effect can lead to a performance degradation of more than 700\%. In
addition, the system does not consider the device-free case.

Different from these systems, the \textit{AROMA} system is unique in
supporting automatic radio map generation for \textbf{both
device-based and device-free} localization systems. To our
knowledge, \textit{AROMA} is the first system to consider radio map
generation for device-free systems and the first to consider human
effects in device-based systems. \textit{AROMA} follows a
site-specific approach combined with the Uniform Theory of
Diffraction (UTD) and multipath fading. In addition \textit{AROMA}
targets high accuracy for the constructed radio maps, which is
needed for localization system. \textit{AROMA}'s performance, in
terms of both predicted signal strength and localization accuracy,
are discussed in Section \ref{sec:validation}.
\section{Conclusion}
\label{sec:conclude}

This paper introduced \textit{AROMA}, a system capable of generating
site-specific radio maps for device-based and device-free
localization systems. \textit{AROMA} combines 3D RF propagation with
human body-scattering effects to achieve high accuracy. We described
the different components of the AROMOA system and how they interact
with each other.

We evaluated the system in two different testbeds. The results show
that \textit{AROMA} can achieve high accuracy with a maximum average
absolute error of 3.18 dBm. We also evaluated the performance of the
system with typical localization systems. Using the radio map
generated by the \textit{AROMA} system, we achieved 1.61m mean error
for the device-based case and 2.44m for the more challenging
device-free case. Our results also showed that the UTD significantly
enhances the performance in the device-free case by more than 700\%.

We also relayed our experience gained while developing the system.
As part of our ongoing work, we are experimenting with different
human models. Using the system for developing new algorithms for
multiple entity tracking and identification for the device-free case
is a possible research area. We are also looking at different
techniques for generating dynamic radio maps and probabilistic radio
maps.

Our experience with the \textit{AROMA} system showed that it
achieves its goals of:
\begin{itemize}
\item High accuracy: through  modeling of the human and combining geometrical optics with the uniform theory of diffraction.
\item Low computational requirements: through the  use of pre-processing, bounding capsules and reception spheres.
\item Minimal user overhead: through automating common tasks, providing default databases for antenna patterns and building materials, and providing a friendly user interface .
\end{itemize}

Moreover, the proposed tool can be used with a wide set of different
applications including site planning for both coverage and
localization, studying new smart APs with dynamic power and channel
adjustment, and designing new localization algorithms, among others.

\vspace{+0.5mm}
\vspace{+2mm}
\bibliographystyle{abbrv}
\scriptsize
\bibliography{ref}

\appendix
\section{RF Propagation}
\label{appendix:A}

Our propagation model uses GO to predict the electric fields at different observation points. It takes into account the electric field characteristics like phase shift and polarization to make the prediction more accurate. The GO is capable of modeling direct, reflected and transmitted electric fields only.  However, the reflection and transmission modeled by GO is for perfect electrically conducting (PEC) surfaces only, while a typical 3D environment consists mainly of non-PEC surfaces. A derivation of an expression for the reflected and transmitted electric fields for a non-PEC surface is provided. We assume that all antennas are vertically polarized and that the emitted waves have spherical wavefronts. Spherical coordinates are used in describing the electric field direction and polarization. Most of the equations in this section are taken from \cite{IntroToUTD}.

\subsection{Electric Field Propagation}

For a spherical wavefront, both radii of curvature are equal ($\rho_1 = \rho_2 = \rho$). The expression describing the propagation of the electric field for a general ray tube at a distance $s$ is
\begin{equation}
E\left(s\right)=E(0)\frac{\rho }{(\rho +s)}e^{-jks}
\label{eq:spherical_electric_field}
\end{equation}

Where\\
\begin{itemize}
	\item $E(0)$ gives the field amplitude, phase and polarization at the reference point $s = 0$.
	\item $s$ is the distance along the ray path from the reference point $s = 0$.
	\item $e^{-jks}$ gives the phase shift along the ray path.
	\item $k$ is the wave number and equals $\frac{2\pi }{\lambda }$
	\item $\rho$ is the principal radius of curvature of the ray tube at the reference point $s = 0$.
\end{itemize}

The radius of curvature at distance $s$ becomes:

\begin{equation}
{\rho }_s=\rho + s
\label{eq:spherical_radius_curvature}
\end{equation}

And the polarization vector remains the same as the reference point.

\subsection{Electric Field Reflection and Transmission for non-PEC surfaces}

Since most surfaces in an indoor environment are non-PEC, a fraction of the electric field incident on the surface reflects off the surface and the rest transmits through it. The ratio of these fractions depends on the characteristics of both mediums as well as the angle of incidence of the incident ray.

Consider a ray tube propagating in the free space from some source. The ray tube impacts a smooth surface at a point $Q_r$. This ray tube is completely described by its central ray vector ${\hat{s}}^i$, principal radii of curvature ${\rho}^i_1$ and ${\rho}^i_2$ at the selected reference point $Q_s$, and its initial electric field is $E^i(Q_s)$ at the reference point $Q_s$.

\begin{figure}
  \centering
  \includegraphics[width=0.25\textwidth]{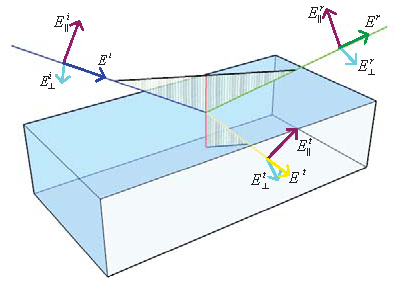}
  \caption{Resolution of electric fields.}
  \label{fig:ef_ref_trans}
\end{figure}

The incident electric field $E^i$ at $Q_r$ can be resolved into two components parallel (${\hat{e}}^i_{\parallel }$) and perpendicular (${\hat{e}}^i_{\bot }$) to the plane of incidence. Similarly, the reflected and transmitted electric fields are resolved into two similar components as shown in Figure \ref{fig:ef_ref_trans}.

The ratio of the amplitudes of the reflected and the transmitted electric fields is given by Fresnel reflection ($\Gamma $) and transmission (${ T}$) coefficients. Fresnel coefficients depend on the polarization of the incident field. For a field polarized in the direction perpendicular to the plane of incidence, Fresnel coefficients are \cite{BornWolf}

\begin{equation}
{\Gamma }_{\bot }{ =}\frac{{{ E}}^{{ r}}_{\bot }}{{{ E}}^{{ i}}_{\bot }}{ =}\frac{{{ n}}_{{ 2}}{\cos  {\theta }^{{ i}}\ }{ -}{{ n}}_{{ 1}}{\cos  {\theta }^{{ t}}\ }}{{{ n}}_{{ 2}}{\cos  {\theta }^{{ i}}\ }{ +}{{ n}}_{{ 1}}{\cos  {\theta }^{{ t}}\ }}
\label{eq:fresnel_ref_perp}
\end{equation}

\begin{equation}
{{ T}}_{\bot }{ =}\frac{{{ E}}^{{ t}}_{\bot }}{{{ E}}^{{ i}}_{\bot }}{ =}\frac{{ 2}{{ n}}_{{ 2}}{\cos  {\theta }^{{ i}}\ }}{{{ n}}_{{ 2}}{\cos  {\theta }^{{ i}}\ }{ +}{{ n}}_{{ 1}}{\cos  {\theta }^{{ t}}\ }}
\label{eq:fresnel_trans_perp}
\end{equation}

While for a field polarized in the direction parallel to the plane of incidence, Fresnel coefficients are :

\begin{equation}
{\Gamma }_{\parallel }{ =}\frac{{{ E}}^{{ r}}_{\parallel }}{{{ E}}^{{ i}}_{\parallel }}{ =}\frac{{ -}{{ n}}_{{ 1}}{\cos  {\theta }^{{ i}}\ }{ +}{{ n}}_{{ 1}}{\cos  {\theta }^{{ t}}\ }}{{{ n}}_{{ 1}}{\cos  {\theta }^{{ i}}\ }{ +}{{ n}}_{{ 2}}{\cos  {\theta }^{{ t}}\ }}
\label{eq:fresnel_ref_par}
\end{equation}

\begin{equation}
{{ T}}_{\bot }{ =}\frac{{{ E}}^{{ t}}_{\bot }}{{{ E}}^{{ i}}_{\bot }}{ =}\frac{{ 2}{{ n}}_{{ 2}}{\cos  {\theta }^{{ i}}\ }}{{{ n}}_{{ 1}}{\cos  {\theta }^{{ i}}\ }{ +}{{ n}}_{{ 2}}{\cos  {\theta }^{{ t}}\ }}
\label{eq:fresnel_trans_par}
\end{equation}

Where
\begin{itemize}
	\item $n_1$ is the absolute refractive index of the first medium.
	\item $n_2$ is the absolute refractive index of the second medium.
	\item $\theta^i $ is the angle of incidence.
	\item $\theta^t $ is the angle of transmission obtained from Snell's Law \ref{eq:snell_law}.
\end{itemize}

The reflected field can be expressed in terms of the incident field and Fresnel reflection coefficients as

\begin{equation}
E^r=\left(E^i\cdot {\hat{e}}^i_{\parallel }\right){\Gamma }_{\parallel }{\hat{e}}^r_{\parallel }+\left(E^i\cdot {\hat{e}}^i_{\bot }\right){\Gamma }_{\bot }{\hat{e}}^r_{\bot }=E^i_{\parallel }{\Gamma }_{\parallel }{\hat{e}}^r_{\parallel }+E^i_{\bot }{\Gamma }_{\bot }{\hat{e}}^r_{\bot }
\label{eq:reflected_incident_field}
\end{equation}

Similarly, the transmitted field can be expressed in terms of the incident field and Fresnel transmission coefficients as

\begin{equation}
E^t=\left(E^i\cdot {\hat{e}}^i_{\parallel }\right){{ T}}_{\parallel }{\hat{e}}^t_{\parallel }+\left(E^i\cdot {\hat{e}}^i_{\bot }\right){{ T}}_{\bot }{\hat{e}}^t_{\bot }=E^i_{\parallel }{{ T}}_{\parallel }{\hat{e}}^t_{\parallel }+E^i_{\bot }{{ T}}_{\bot }{\hat{e}}^t_{\bot }
\label{eq:transmitted_incident_field}
\end{equation}

A negative reflection coefficient means an increase in the phase shift by $180^\circ$.

By the summation of the two components using vector sum, the amplitude and the polarization of the resultant the reflected or the transmitted field is obtained.

The incident ray tube is spherical, ${\rho }^i_1={\rho }^i_2=\rho $, and all surfaces are assumed planar. The radii of curvature of the reflected and transmitted ray tubes then reduce to

\begin{equation}
	{\rho }^{r,t}_1={\rho }^{r,t}_2=\rho_s
\label{eq:ref_trans_rad_cur}
\end{equation}

\subsection{Wedge Diffraction}
\begin{figure}
    \centering
        \includegraphics[width=0.25\textwidth]{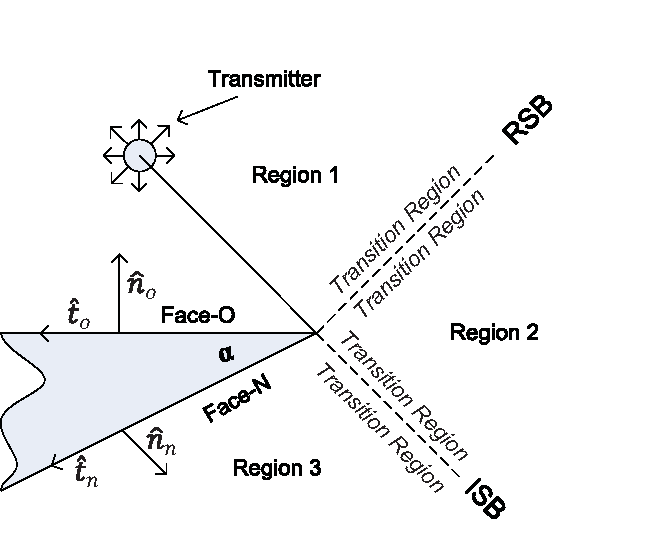}
    \caption{Shadow regions.}
    \label{fig:shadow_regions}
\end{figure}

A shadow region is the area that a GO ray can't reach due to the presence of obstacles in its path. A shadow boundary is a boundary that defines the shadow region. \ref{fig:shadow_regions} shows that there exist shadow regions for direct and reflected rays. GO is not sufficient to model these regions since it predicts Zero electric field in the shadow regions. Our model uses the Uniform Theory of Diffraction (UTD) to model wedge diffraction because of its ability to accurately predict the electric field in the shadow regions as well as its ability to give a non singular solution at the transition regions.

Figure \ref{fig:shadow_regions} shows the geometry of a wedge. A wedge consists of two faces, O-face and N-face, and an edge. Each face has unit tangent and normal vectors. Tangents are directed away from the edge while normals are directed towards the outside of the wedge. The inner angle of the wedge is $\alpha$.

The plane of incidence is defined by ${\hat{s}}^{'}$ and $\hat{e}$, while the plane of diffraction is defined by $\hat{s}$ and $\hat{e}$.

The incident electric field is resolved into two components parallel (${\widehat{\beta }}^{'}_0$) and perpendicular (${\hat{\varphi}}^{'}$) to the plane of incidence. The diffracted electric field is also resolved into two components parallel ($\widehat{\beta }$) and perpendicular ($\hat{\varphi}$) to the plane of diffraction.

\subsubsection{Diffracted Electric Field}

\begin{figure}
    \centering
        \includegraphics[width=0.3\textwidth]{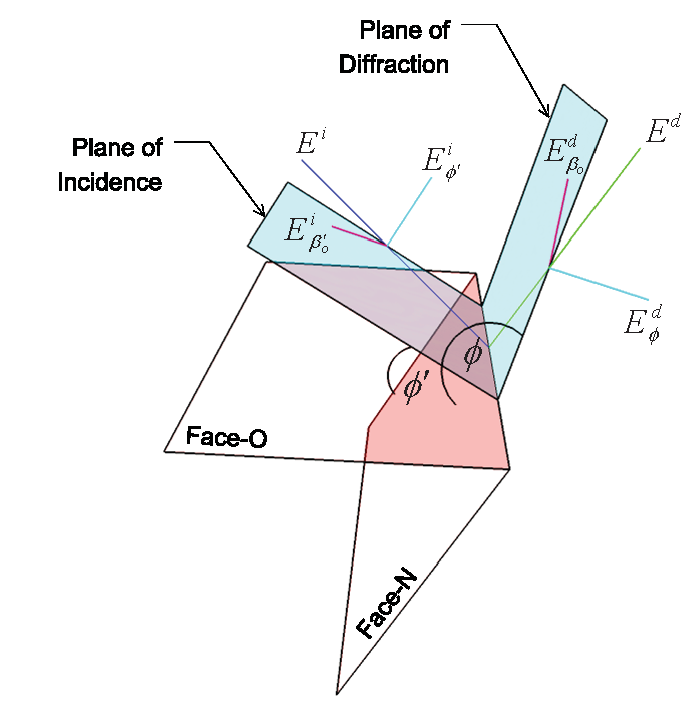}
    \caption{Resolution of electric fields.}
    \label{fig:ef_dif}
\end{figure}

Using figure \ref{fig:ef_dif}, the incident electric field can be expressed as

\begin{equation}
E^i=E^i_{{\beta }^{'}_o}{\widehat{\beta }}^{'}_0+E^i_{\phi^{'}}{\hat{\phi}}^{'}
\label{eq:w11}
\end{equation}

and the diffracted electric field can be expressed as

\begin{equation}
E^d=E^d_{{\beta }_0}{\widehat{\beta }}_0+E^d_\phi\hat{\phi}
\label{eq:w12}
\end{equation}

The diffracted electric field at a distance $s$ on the path of the diffracted ray is given by

\begin{equation}
E^d=D.E^i.A\left(s\right)e^{-jks}
\label{eq:w15}
\end{equation}

Where :
\begin{itemize}
	\item D is the UTD diffraction coefficient.
		\begin{equation}
			D=\left[ \begin{array}{cc}
			-D_s & 0 \\ 
			0 & -D_h \end{array}
			\right]
			\label{eq:w16}
		\end{equation}
	\item A(s) is the spreading factor defined by:
		\begin{equation}
		A\left(s\right)=\sqrt{\frac{\rho }{s\left(\rho +s\right)}}
		\label{eq:w17}
		\end{equation}
	\item $\rho $ is the radius of curvature of the incident ray tube at the point of intersection with the edge.
\end{itemize}

\subsubsection{UTD Diffraction Coefficients for PEC wedges}

\begin{equation}
D_{s,h}\left(L^i,L^{ro},L^{rn},\phi,\phi^{'},{\beta }_0,n\right)=D_1+D_2+R_{s,h}(D_3+D_4)
\label{eq:w18}
\end{equation}

Where $R_{s,h}$ are the reflection coefficients of the wedge at the edge. For a PEC surface $R_s=-1$ and $R_h=1$.

The components of the diffraction coefficients are given by

\begin{equation}
D_1=\frac{-e^{-j\frac{\pi }{4}}}{2n\sqrt{2\pi k}{\sin  {\beta }_0\ }}{\cot  \left[\frac{\pi +(\phi-\phi^{'})}{2n}\right]F\left[kL^ia^+(\phi-\phi^{'})\right]\ }
\label{eq:w19}
\end{equation}

\begin{equation}
D_2=\frac{-e^{-j\frac{\pi }{4}}}{2n\sqrt{2\pi k}{\sin  {\beta }_0\ }}{\cot  \left[\frac{\pi -(\phi-\phi^{'})}{2n}\right]F\left[kL^ia^-(\phi-\phi^{'})\right]\ }
\label{eq:w20}
\end{equation}

\begin{equation}
D_3=\frac{-e^{-j\frac{\pi }{4}}}{2n\sqrt{2\pi k}{\sin  {\beta }_0\ }}{\cot  \left[\frac{\pi +(\phi+\phi^{'})}{2n}\right]F\left[kL^{rn}a^+(\phi+\phi^{'})\right]\ }
\label{eq:w21}
\end{equation}

\begin{equation}
D_4=\frac{-e^{-j\frac{\pi }{4}}}{2n\sqrt{2\pi k}{\sin  {\beta }_0\ }}{\cot  \left[\frac{\pi -(\phi+\phi^{'})}{2n}\right]F\left[kL^{ro}a^-(\phi+\phi^{'})\right]\ }
\label{eq:w22}
\end{equation}

The functions a$\pm$ are defined as:

\begin{equation}
a^{\pm }\left({\beta }^{\pm }\right)=2{{\cos }^{{ 2}} \left(\frac{2n\pi N^{\pm }-{\beta }^{\pm }}{2}\right)\ }
\label{eq:w23}
\end{equation}

The meaning and values of the parameters included can be obtained from \cite{IntroToUTD}.

\subsubsection{UTD Diffraction Coefficients for non-PEC wedges}

A number of heuristics were made to derive an expression for the diffraction coefficients for non-PEC surfaces. One of these heuristics uses Fresnel reflection and transmission coefficients as follows \cite{SiteSpecificThesis}

\begin{equation}
D_s\left(L^i,L^{ro},L^{rn},\phi,\phi^{'},{\beta }_0,n\right)=D_1+D_2+{\Gamma }^{{ n}}_{\parallel }D_3+{\Gamma }^{{ o}}_{\parallel }D_4
\label{eq:w29}
\end{equation}

\begin{equation}
D_h\left(L^i,L^{ro},L^{rn},\phi,\phi^{'},{\beta }_0,n\right)=D_1+D_2+{\Gamma }^{{ n}}_{\bot }D_3+{\Gamma }^{{ o}}_{\bot }D_4
\label{eq:w30}
\end{equation}

Where
\begin{itemize}
	\item ${\Gamma }^{{ o}}_{\parallel }$ and ${\Gamma }^{{ n}}_{\parallel }$ are Fresnel's reflection coefficients for parallel polarization of the o-Face and n-Face respectively.
	\item ${\Gamma }^{{ o}}_{\bot }$ and ${\Gamma }^{{ n}}_{\bot }$ are Fresnel's reflection coefficients for perpendicular polarization of the o-Face and n-Face respectively.	
\end{itemize}

\section{Human Modeling}
\label{appendix:B}

In order to predict the human body-scattering effects in the indoor environment, a UTD-based propagation model is used. In the model, the human body is approximated with a perfect conducting circular cylinder\cite{UTDTalbi} of  radius 0.20 m, and height of 2 m. At microwave frequencies and higher, the human body constitutes a perfectly conducting and impassible reflector for electromagnetic waves \cite{UTDTalbi}.

\subsection{Geometrical Configuration}

An incident ray on a cylindrical surface may be either reflected or diffracted, according to the location of intersection between the ray and the cylindrical surface. A ray that falls tangential to the cylinder grazes the surface in the same plane of incidence. An extension of the incident ray beyond the point of grazing at the convex surface of the cylinder defines the Shadow Boundary (SB), which splits the space outside the surface into the lit and shadow regions \cite{IntroToUTD}.

\begin{figure}
    \centering
        \includegraphics[width=0.30\textwidth]{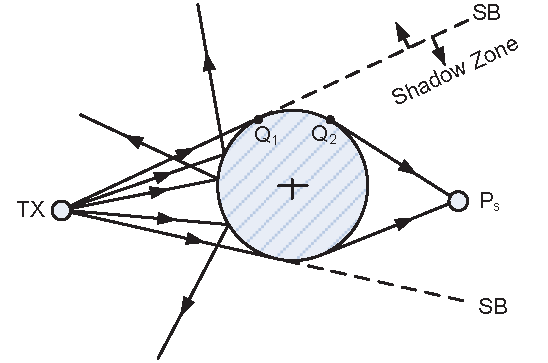}
    \caption{Geometrical configuration of the human shadowing problem.}
    \label{fig:creeping}
\end{figure}

Regions in the vicinity of the shadow boundary are called the transition regions. It is at the transition regions where the GO and GTD model fails as the incident field drops to zero, and the predicted field becomes infinite, and thus UTD was introduced \cite{IntroToUTD}. Reflected rays from the cylinder falls in the lit region, while grazing rays detach from the surface into diffracted rays and fall in the shadow zone, which is the non-illuminated exterior region of the cylinder. A grazing ray path is determined by the shortest distance from the ray source to an observation point $P_s$, and creeping on the cylindrical surface \cite{IntroToUTD}.

\subsection{Reflection Off a Cylindrical Surface}

The UTD expression of the field reflected off a cylindrical surface from an incident electric field is given by \cite{IntroToUTD}:

\begin{equation}
\label{eq:UTDReflectedField} 
E^r\left(p\right)=E^i\left(Q_r\right)\cdot R_{s,h}\cdot A\left(s\right)\cdot e^{-jks^r}
\end{equation}

Where:
\begin{itemize}
	\item $E^i\left(Q_r\right)$ is the electric field incident to the cylinder at $Q_r$.
	\item $R_{s,h}$ are the soft and hard UTD reflection coefficients.
	\item $A\left(s\right)$ is the 3-D spreading factor of reflection.
	\item $e^{-jks^r}$ is the phase shift of the reflected field.
	\item $s^r$ is the distance between reflection point and observation point.
\end{itemize}

The soft and hard reflection coefficients are calculated from the following expression \cite{IntroToUTD}:

\begin{equation}
\label{eq:UTDReflectionCoefficient} 
R_{s,h}=-\sqrt{\frac{-4}{\xi }}\cdot e^{-j\pi /4}\cdot e^{-j{\xi }^3/12}\left[\frac{-F(X_p)}{2\xi \sqrt{\pi }}+\left\{ \begin{array}{c}
p^*(\xi ) \\ 
q^*(\xi ) \end{array}
\right\}\right]
\end{equation}

Where:
\begin{itemize}
	\item $p^*(\xi )$ and $q^*\left(\xi \right)$ are the soft and hard Fock scattering functions.
	\item $F(X_p)$ is the UTD reflection transition function which ensures that fields surrounding the transition regions remains bounded.
	\item $X_p{ (}\ge 0)$ is the transition function parameter, which is defined from \cite{UTDSmoothPathak}:
\end{itemize}

\begin{equation}
\label{eq:ReflectionTransitionParam}
X_p = 2kL_p{\cos}^2\left({\theta }^i\right)
\end{equation}
Where $\xi (\le 0)$ is the Fock parameter. It is defined from \cite{UTDSmoothPathak}:

\begin{equation}
\label{eq:ReflectionFockParameter}
\xi = -2m\left(Q_r\right){ \cos}({\theta }^i)
\end{equation}
Where $m\left(Q_r\right)$ is the curvature parameter at the reflection point. It is calculated as \cite{IntroToUTD}:

\begin{equation}
m\left(Q_r\right)={\left[\frac{ka_t(Q_r)}{2}\right]}^{1/3}
\label{eq:}
\end{equation}

Where $a_t(Q_r)$ is the radius of curvature of the surface in the plane of incidence at $Q_r$. Computation-friendly expressions for the transition and the scattering functions can be found in \cite{IntroToUTD}, and \cite{FockRef}.

$L_p$ is the 3-D distance parameter, which is calculated from \cite{UTDSmoothPathak}:

\begin{equation}
\label{eq:3DReflectionDistanceParam} 
L_p=\frac{{\rho }^i_1{\rho }^i_2}{\left({\rho }^i_1+s^r\right)\left({\rho }^i_2+s^r\right)}\cdot \frac{s^r\left({\rho }^r_2+s^r\right)}{{\rho }^r_2}
\end{equation}

Where:
\begin{itemize}
	\item ${\rho }^i_1$ (${\rho }^i_2$) is the radius of curvature of the incident wave in (transverse to) the plane of incidence.
	\item ${\rho }^r_2$ is the radius of curvature of the reflected wave transverse to the plane of reflection. 
\end{itemize}

In our implementation, waves are assumed to be spherical. In that case, ${\rho }^i_1={\rho }^i_2=s^i$ where $s^i$ is the total unfolded distance from the transmitter to the point of reflection.  The reflected wavefront radii of curvature are calculated from \cite{IntroToUTD}:

\begin{equation}
\label{eq:ReflectedRadii} 
\frac{1}{{\rho }^r_{1,2}}=\frac{1}{s^i}+\frac{1}{f_{1,2}}
\end{equation}

Where:\\
\begin{multline}
\label{eq:F12ReflectedRadii}
f_{1,2}=\frac{1}{\cos(\theta^i)} \left[ \frac{\sin^2(\theta_2)}{a_1} + \frac{\sin^2(\theta_1)}{a_2} \right]
\\
\pm {\left[ \frac{1}{\cos^2(\theta^i)} \left( \frac{\sin^2(\theta_2)}{a_1} + \frac{\sin^2(\theta_1)}{a_2} \right) - \frac{4}{a_1 a_2} \right]}^{1/2}
\end{multline}

The terms $a_1$and $a_2$ are the principal radii of curvature of the surface at the incidence point. They measure how the surface bends by different amounts in different directions at a given point. In the case of a circular cylinder, $a_1=radius$ and $a_2=infinity$, hence the previous expression reduces to:

\begin{equation}
\label{eq:F12ReflectedRadiiSpherical} 
f_{1,2}=\frac{1}{{\cos}({\theta }^i)}\left[\frac{{{ \sin}}^{{ 2}}({\theta }_2)}{a_1}\right]\pm {\left[\frac{1}{{{\cos}}^{{ 2}}({\theta }^i)}\left(\frac{{{ \sin}}^{{ 2}}({\theta }_2)}{a_1}\right)\right]}^{1/2}
\end{equation}
\\
The quantities ${{ \sin}}^{{ 2}}({\theta }_1)$ and ${{ \sin}}^{{ 2}}({\theta }_2)$ are calculated from \cite{IntroToUTD}:

\begin{equation}
\label{eq:Sin2Alpha} 
{{\sin }^{{ 2}} \left({\theta }_1\right)\ }={{\cos }^{{ 2}} \left(\alpha \right)\ }+{{ \sin}}^{{ 2}}(\alpha ){ \ }{{ \cos}}^{{ 2}}(\theta^i)
\end{equation}

\begin{equation}
\label{eq:Cos2Alpha} 
{{\sin }^{{ 2}} \left({\theta }_2\right)\ }={{\sin }^{{ 2}} \left(\alpha \right)\ }+{{ \cos}}^{{ 2}}(\alpha ){ \ }{{ \cos}}^{{ 2}}(\theta^i) 
\end{equation}

\begin{figure}
    \centering
        \includegraphics[width=0.3\textwidth]{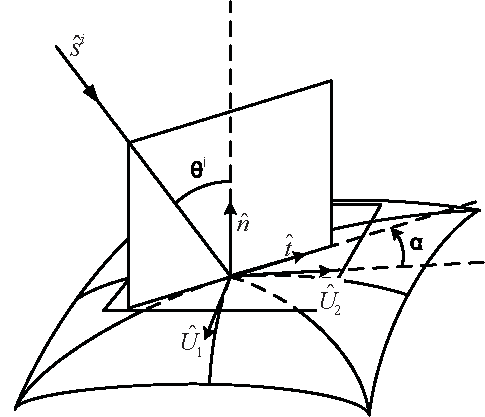}
    \caption{Principal directions at the point of incidence.}
    \label{fig:principal_directions}
\end{figure}

Where $\alpha$ is the angle between the principal plane of the surface at the point of incidence and the plane of incidence as shown in Figure \ref{fig:principal_directions}.

The value of $\alpha$ is calculated from \cite{UTDSmoothPathak}:

\begin{equation}
\label{eq:Alphat1} 
t_1=-{\hat{s}}^i\cdot {\hat{U}}_1
\end{equation}

\begin{equation}
\label{eq:Alphat2} 
t_2=-{\hat{s}}^i\cdot {\hat{U}}_2
\end{equation}

\begin{equation}
\label{eq:Alpha} 
\alpha =\left|{\tan^{-1} \frac{t_1}{t_2}\ }\right|
\end{equation}

Where ${\hat{U}}_1$ and ${\hat{U}}_2$ are the principal directions of the surface at the point of incidence. In the case of a cylindrical surface, ${\hat{U}}_2$ is parallel to the cylinder generator, and ${\hat{U}}_1$ is perpendicular to ${\hat{U}}_2$.

Finally, the reflected spherical wave front radii are calculated as:

\begin{equation}
\label{eq:ReflectedSphericalRadii}
{\rho }^r_{1,2}=\frac{\rho^r_1+\rho^r_2}{2}
\end{equation}

When an incident field reflects from a cylindrical surface, the reflected field amplitude varies from that of the incident field. The 3-D spreading factor governs the amplitude variation of the reflected field in 3D. It is calculated from \cite{UTDSmoothPathak}:

\begin{equation}
\label{eq:3dSpreadingFactor}
A\left(s\right)=\sqrt{\frac{{\rho }^r_1{\rho }^r_2}{\left({\rho }^r_1+s^r\right)\left({\rho }^r_2+s^r\right)}}
\end{equation}

\subsection{Diffraction Off a Cylindrical Surface}

An incident tangential ray at $Q_1$ creeps the cylinder surface and detaches at $Q_2$. The location of $Q_2$ depends on the location of the observation point so that the total distance from the point of incidence to the observation point, including the creeping distance, is minimal. 

The field diffracted from a cylindrical surface is given by \cite{UTDSmoothPathak}:

\begin{equation}
\label{eq:UTDDiffractedField}
E^d\left(p\right)=E^i\left(Q_1\right)\cdot T_{s,h}\cdot EC\cdot A\left(s\right)\cdot e^{-jks^r}
\end{equation}

Where:
\begin{itemize}
	\item $E^i\left(Q_1\right)$ is the incident field on the cylinder at $Q_1$.
	\item $T_{s,h}$ are the soft and hard UTD diffraction coefficient.
	\item $A\left(s\right)=\sqrt{\frac{\rho^r_2}{s^r\left(\rho^r_2+s^r\right)}}$ is the 3-D spreading factor of diffraction.
	\item $EC=\sqrt{\frac{\rho^i_2}{\rho^i_2+t}}$ is the conservation of energy term in the case of spherical waves, where $t$ is the creep distance along the surface from $Q_1$ to $Q_2$.
\end{itemize}

The soft and hard UTD diffraction coefficients are given by \cite{UTDSmoothPathak}:

\begin{multline}
\label{eq:UTDDiffractionCoefficient}
T_{s,h} = -\sqrt{m(Q_1) m(Q_2)} \cdot \sqrt{\frac{2}{k}} \cdot e^{-j \pi /4} \cdot e^{-jk} 
\\
\cdot \left[
\frac{-F(X_d)}{2 \xi \sqrt{\pi}} + \left\{ \begin{array}{c}
p^*(\xi ) \\ 
q^*(\xi ) \end{array}
\right\}\right]
\end{multline}

Where:
\begin{itemize}
	\item $F(X_d)$ is the diffraction transition function.
	\item $X_d=\frac{k L_d \xi^2}{2m\left(Q_1\right)m(Q_2)}$ is the transition function argument.
	\item $L_d$ is the 3-D distance parameter.
	\item $\xi =\int_t{\frac{m(Q_t)}{a_t}}$ is the Fock parameter.
\end{itemize}

Transition and Fock functions are computed as described in the previous section.

The two radii of curvature of the diffracted wave at the observation point are calculated from \cite{UTDSmoothPathak}:

\begin{equation}
\label{eq:DiffractedRadii}
\rho^r_1 = \rho^i_b + s^r + t \rho^r_2 = s^r
\end{equation}

Where:
\begin{itemize}
	\item $t$ is the creeping distance.
  \item $s^r$ is the distance between the observation point and the detachment point $Q_2$.
	\item $\rho^i_b$ is the radius of curvature of the incident wave in the plane transverse to the incident plane, which is calculated from \cite{UTDSmoothPathak}:	
\end{itemize}

\begin{equation}
\label{eq:RooTransverse}
\frac{1}{\rho^i_b}=\frac{\sin^2(\alpha)}{\rho^i_1}+\frac{\cos^2(\alpha)}{\rho^i_2}
\end{equation}

As we assume spherical waves, that is $\rho^i_1 = \rho^i_2 = s^i$, the previous equation reduces to: 

\begin{equation}
\rho^i_b = \rho^i_1 = \rho^i_2
\label{eq:}
\end{equation}

The final radii of curvature of the diffracted ray are then calculated as the average of its two radii of curvature.

\end{document}